\documentclass[11pt]{JHEP3}
\usepackage{amsfonts,amssymb,amsmath}
\usepackage[dvips]{graphicx}
\usepackage{psfrag}
\usepackage{epsfig}

\newcommand{\NCS}{N_{\rm CS}}
\newcommand{\nw}{N_{\rm w}}
\newcommand{\nnw}{n_{\rm w}}
\newcommand{\nncs}{n_{\rm CS}}

\newcommand{\mueff}{\mu_{\rm eff}}
\newcommand{\beaa}{\begin{eqnarray*}}
\newcommand{\eeaa}{\end{eqnarray*}}
\newcommand{\bc}{\begin{center}}
\newcommand{\ec}{\end{center}}

\newcommand{\intvecx}{\int d^3 x\,}
\newcommand{\intvecy}{\int d^3 y\,}

\newcommand{\veck}{{\bf k}}

\newcommand{\vecx}{{\bf x}}       
\newcommand{\vecy}{{\bf y}}

\newcommand{\vecr}{{\bf r}}

\newcommand{\al}{\alpha}
\newcommand{\bt}{\beta}
\newcommand{\gm}{\gamma}
\newcommand{\dl}{\delta}
\newcommand{\ep}{\epsilon}

\newcommand{\kp}{\kappa}
\newcommand{\lm}{\lambda}
\newcommand{\rh}{\rho}
\newcommand{\sg}{\sigma}

\newcommand{\ph}{\phi}
\newcommand{\vr}{\varphi}
\newcommand{\ch}{\chi}

\newcommand{\om}{\omega}
\newcommand{\Om}{\Omega}

\newcommand{\Ph}{\Phi}
\newcommand{\Phd}{\Phi^{\dagger}}

\newcommand{\half}{\frac{1}{2}}
\newcommand{\quart}{\frac{1}{4}}

\newcommand{\Tr}{\mbox{Tr}}

\newcommand{\dmu}{\partial_{\mu}}

\newcommand{\eela}[1]{\label{#1}\end{equation}}
\newcommand{\eeala}[1]{\label{#1}\end{eqnarray}}
\newcommand{\be}{\begin{equation}}
\newcommand{\ee}{\end{equation}}
\newcommand{\bea}{\begin{eqnarray}}
\newcommand{\eea}{\end{eqnarray}}

\newcommand{\mh}{m_{\rm H}}

\title{Chern-Simons and winding number in a tachyonic electroweak
  transition}

\author{Meindert van der Meulen\\
Institute for Theoretical Physics, University of Amsterdam\\
Valckenierstraat 65, 1018 XE Amsterdam, The Netherlands\\
E-mail: \email{mmeulen@science.uva.nl}}
\author{D\'enes Sexty\\
Department of Atomic Physics\\
E\"otv\"os University, Budapest, Hungary\\
E-mail: \email{denes@achilles.elte.hu}}
\author{Jan Smit\\
Institute for Theoretical Physics, University of Amsterdam\\
Valckenierstraat 65, 1018 XE Amsterdam, The Netherlands\\
E-mail: \email{jsmit@science.uva.nl}}
\author{Anders Tranberg\\
Department of Physics and Astronomy, University of Sussex\\
Falmer, Brighton, East Sussex BN1 9QH. UK\\
E-mail: \email{a.tranberg@sussex.ac.uk}}
\preprint{ITFA-2005-46}
\abstract{ We investigate the development of winding number and
Chern-Simons number in a tachyonic transition in the SU(2) Higgs
model, motivated by the scenario of cold electroweak baryogenesis. We
find that localized configurations with approximately half-integer
winding number, dubbed half-knots, play an important role in this
process.  When the Chern-Simons number adjusts locally to the winding
number, the half-knots can stabilize and acquire half-integer
Chern-Simons number as well. We present two examples from numerical
simulations: one half-knot that stabilizes early and one that gives
rise to a late sphaleron transition.  We also study the winding number
distribution after the transition and present new results on the
development of the Chern-Simons susceptibility. }

\begin{document}
\section{Introduction}
Baryogenesis, the creation of the baryon asymmetry in the universe, is
a long standing problem in cosmology.  It dates back to 1967, when
Sakharov suggested that the baryon asymmetry is not an initial
condition of the universe, but might be created later in a process
based on particle physics \cite{Sakharov:1967dj}.  This idea has
gained support from the inflationary scenario, since inflation is
supposed to have diluted any pre-existing asymmetry. Sakharov
formulated his well-known conditions for baryogenesis: baryon number
conservation, C, and CP must be violated, and a state of
non-equilibrium must prevail.

Of the many particle physics scenarios that have been proposed in the
past decades implementing such a process, electroweak baryogenesis
\cite{Cohen:1993nk,Rubakov:1996vz,Riotto:1999yt} is interesting in
that it suggested the possibility to explain the baryon asymmetry
using mostly Standard Model physics. In this scenario the baryon
number violation is caused by the anomaly that relates a change in
baryon number $B$ to a change in Chern-Simons number $N_{\rm CS}$ of
the electroweak gauge fields:
\begin{equation}
\Delta B = 3 \langle \Delta N_{\rm CS}\rangle. \label{anomaly}
\end{equation}
Furthermore the Standard Model violates C, it has a CP violating phase
in the CKM quark mixing matrix, and the out-of-equilibrium conditions
can be provided by an electroweak phase transition. This phase
transition was supposed to be caused by the lowering temperature of
the universe, and to be sufficiently out of equilibrium it had to be
of first order.  However subsequent work has shown that for the
experimentally allowed range of the Higgs mass, the electroweak phase
transition is only a crossover (see e.g.\ \cite{Kajantie:1996kf}). It
is widely believed that a crossover transition is too close to
equilibrium for creation of the asymmetry.  Furthermore, the CKM
CP violation has been found to be much too small
\cite{Shaposhnikov:1987tw,Gavela:1993ts,Huet:1994jb}.

A few years ago, new scenarios were proposed
\cite{Garcia-Bellido:1999sv,Krauss:1999ng}, in which electroweak
baryogenesis takes place during a {\em tachyonic transition}.  In such
a transition the effective mass term in the Higgs potential starts
being positive, and can change sign due to the coupling to an inflaton
field, as in hybrid inflation \cite{Linde:1993cn}. The accompanying
instability can lead to strongly out-of-equilibrium conditions with
large occupation numbers in the Higgs and gauge fields, during which
the energy in the Higgs field is transferred to the other fields by
wave-like `rescattering'.  The process is called tachyonic preheating
\cite{Felder:2000hj}.  During the transition there can be substantial
changes in the Chern-Simons number, and also the baryon number via the
anomaly equation (\ref{anomaly}). The universe was assumed to be cold
after electroweak-scale inflation, so initially the transition takes
place at practically zero temperature.

In subsequent papers the scenario was further refined and tested.
Considerations of quantum corrections led to a change of model to
inverted hybrid inflation \cite{Copeland:2001qw}, in which the
inflaton rolls away from the origin instead of towards it. In
\cite{vanTent:2004rc} it was shown how WMAP data constrain the
parameters of a model and it was noted that it might be falsified by
the LHC.  The transition was studied by analytic and numerical methods
\cite{Garcia-Bellido:2002aj,Smit:2002yg,Skullerud:2003ki,Garcia-Bellido:2003wd,Tranberg:2003gi},
and the magnitude of the asymmetry generated by a form of CP violation
was computed in \cite{Smit:2002yg,Tranberg:2003gi}.

The CP violating term in the Lagrangian that was used in
\cite{Tranberg:2003gi} does not occur in the Standard Model.  Of
course, one is also interested in the CP violation from the CKM
matrix. As mentioned above, this CP violation has been estimated to be
much too small for baryogenesis \cite{Shaposhnikov:1987tw,
Gavela:1993ts,Huet:1994jb}, but these estimates do not seem to apply
to a tachyonic transition at zero temperature. In fact, it has been
suggested \cite{Smit:2004kh} that the effect might be much larger in
this case.  It is therefore important to make sure whether the CP
violation of the Standard Model is sufficient to produce the baryon
asymmetry.

Trying to investigate this problem by numerical simulation with
three generations of fermions is a practically impossible
task. Instead we have in mind a more tractable approach: if the
changes of $N_{\rm CS}$ occur in a certain type of local field
configuration, we could estimate the produced asymmetry by simulating
only this local configuration. There is reason to believe that the
change of $N_{\rm CS}$ indeed occurs in local configurations: in
\cite{Garcia-Bellido:2003wd} evidence is found for local structures in
numerical simulations, and in \cite{Krauss:1999ng,Copeland:2001qw} it
is suggested that topological defects called textures play a role in
this process. The presence of a texture depends on the winding number
of the Higgs field $N_{\rm w}$. In the vacuum $N_{\rm CS} = N_{\rm
w}$. A texture is a configuration which has winding number different
from the Chern-Simons number. It is unstable and can decay either by
changing the winding number or the Chern-Simons number. In
\cite{Turok:1990in,Turok:1990zg} a scenario for electroweak
baryogenesis is investigated in which the change of Chern-Simons
number occurs during the decays of textures. The textures were
supposed to be formed directly after a first order electroweak phase
transition, and the produced asymmetry was estimated by simulating a
{\em single} texture and its decays. Such an approach was investigated
further in ref.\ \cite{Lue:1996pr}, where it was concluded that it is
unlikely to be successful and that the asymmetry depends on too many
variables to bypass a fully-fledged numerical simulation.  We believe
this conclusion is not so clear cut and consider it worthwhile to
understand more fully the mechanism that changes Chern-Simons numbers
in tachyonic transitions.

In this paper we study the production of winding and Chern-Simons
number in a tachyonic transition.  We shall argue that instead of
textures, related configurations with half-integer winding number are
important.  We call such configurations half-knots.\footnote{The word
half-knot appeared earlier in \cite{Turok:1989ai}.}  These typically
occur in regions where the Higgs magnitude has a small minimum.  They
can be stabilized when the Chern-Simons number density adjusts to the
winding number density and the Higgs field relaxes towards its ground
state, leaving a blob-like half-knot both in winding number and in
Chern-Simons number.  Half-knots have a rather high winding number
density in their center and can be visualized in numerical
simulations. We present some examples in detail.

In section \ref{swsu2} we review the Chern-Simons number, winding
number and winding configurations in this model. Next we turn to the
tachyonic transition and discuss our expectations with respect to the
half-knots in this transition, in section \ref{stt}. In section
\ref{sns} we present the results of the numerical simulations, and we
discuss the results in section \ref{sd}.

\section{Winding in the SU(2) Higgs model} \label{swsu2}
In this section we review some topological features of the SU(2) Higgs
model, since it is the part of the Standard Model that plays a
dominant role in the tachyonic transition.  First we introduce the
model and define the Higgs winding number and the Chern-Simons
number. Then we discuss topological defects that may play a role in
the transition: textures, sphalerons and half-knots.

\subsection{SU(2) Higgs model}
The action is given by
\begin{equation}
S=-\int d^4\!x \left[\frac{1}{2 g^2} \mbox{Tr} F_{\mu\nu}
F^{\mu\nu} +
\half\,
\mbox{Tr} \left[ \left( D_{\mu} \Phi \right)^{\dagger}
D^{\mu} \Phi \right] + \lambda \left(
\half\,
\mbox{Tr} \left[\Phi^{\dagger}
\Phi \right] - \frac{v^2}{2} \right)^2 \right], \label{action}
\end{equation}
where the field strength is $F_{\mu\nu} = \partial_{\mu} A_{\nu} -
\partial_{\nu} A_{\mu} -i [A_{\mu}, A_{\nu} ]$, the vector potential
can be written as $A_{\mu} = A_{\mu}^a \tau^a/2$, and the covariant
derivative acting on the Higgs field is $D_{\mu} \Phi =
(\partial_{\mu} - iA_{\mu} ) \Phi$. We use a metric with signature
(-1,1,1,1) and for the Higgs field we use the matrix notation:
\begin{equation}
\Ph =
\begin{pmatrix}
\vr_d^* & \vr_u \\ -\vr_u^* &\vr_d
\end{pmatrix} =
\frac{\rho}{\sqrt{2}} \;
\Omega,
\quad
\rh^2 = 2(\vr_u^*\vr_u + \vr_d^*\vr_d),
\quad
\Omega(x) \in \mbox{SU(2)}. \label{higgsdef}
\end{equation}
We call $\rho$ the Higgs length. The Higgs and W masses are
given by $m_{\rm H}=\sqrt{2 \lambda} \; v$ and $m_W= gv/2$,
we also use the notation
\be
\mu = \sqrt{\lm v^2}.
\ee

As part of an extended theory, the mass term $-\lm v^2 \half\, \Tr\,
\Phd\Ph$ is to be replaced by an effective mass term 
\be
\mueff^2\half\,\Tr\, \Phd\Ph,
\label{mueffintro}
\ee 
where $\mueff^2$ depends on time through the coupling to another
field (inflaton).  Initially it is positive, and when it changes to
negative the tachyonic transition starts. Eventually $\mueff^2$ will
relax to the Standard Model value 
\be 
\mueff^2 \to -\lm v^2 = -\mu^2.
\ee 
The rate of change of $\mueff$ depends on further details of the
theory.

Throughout this paper we we use the so-called `temporal gauge'
$A_0=0$, which still leaves the freedom to do time-independent gauge
transformations.

\subsection{Topology in the SU(2) Higgs model}
The non-conservation of baryon number in the Standard Model
follows from the anomaly in the divergence of the baryon current,
\bea
\dmu j^\mu_B &=& 3\, q,
\label{anomdiv}
\\
q &=& \frac{1}{32\pi^2}\, \ep^{\kp\lm\mu\nu} \Tr\, F_{\kp\lm} F_{\mu\nu}
= \dmu j^\mu_{\rm CS}.
\label{qdef}
\eea
Here 3 is the number of generations and $j^\mu_{\rm CS}$ is the Chern-Simons
current; $q$ is sometimes called the topological charge density, since, for
classical fields, and in euclidean space-time, its integral over a
four-dimensional manifold without boundary
is an integer, the topological charge.
Taking the expectation value of (\ref{anomdiv}) in the initial
(Heisenberg) state and
integrating between (real) times 0 and $t$ gives
\be
B(t) - B(0) = \int_0^t dx^0\intvecx \langle 3 q\rangle =
3\langle \NCS(t) - \NCS(0)\rangle,
\label{intanomdiv}
\ee
with $B = \langle\intvecx J^0_B\rangle$ the baryon number and
\be
\NCS = \intvecx j^0_{\rm CS},
\quad
j^0_{\rm CS} =
-\frac{1}{16 \pi^2} \epsilon^{jkl} \mbox{Tr}
  \left[ A_j \left( F_{kl}+i \frac{2}{3} A_k A_l\right) \right],
\label{ncs}
\ee
the Chern-Simons number operator.
We assumed that $\intvecx \partial_k \langle j^k_{\rm CS}\rangle$
vanishes, e.g.\ in a model with periodic boundary conditions, or
because the fields vanish sufficiently fast at spatial infinity.

At this stage the Chern-Simons number and current are
still operators, whereas the baryon number $B$ is a c-number in the way
we have written it. In the following we shall make a classical approximation
(see section \ref{stt}),
and therefore we assume from now on that all fields are
classical. Note that $\NCS(t)$ and $\NCS(t)-\NCS(0)$ are generically
not integers. 

The winding number of the Higgs field is given by
\begin{align}
N_{\rm w} &=
\intvecx \nnw,
\label{nw}
\\
\nnw &=
\frac{1}{24 \pi^2}\,  \epsilon^{ijk}
\mbox{Tr} \left[ \partial_i \Omega \Omega^{-1} \partial_j \Omega
\Omega^{-1} \partial_k \Omega \Omega^{-1} \right],
\label{nnw}
\end{align}
where $\Om$ is given implicitly by (\ref{higgsdef}); this is a valid
definition as long as $\rho \neq 0$ everywhere. Classical vacuum
configurations are given by
\begin{equation}
\Phi =
\frac{v}{\sqrt{2}}\,
\Omega, \hspace{1cm} A_j = -i
\partial_j \Omega
\Omega^{-1}, \hspace{1cm} \Omega \in
\mbox{SU(2)}
\label{vac}.
\end{equation}
Here $\Om$ is arbitrary. It is easy to check that in the vacuum
\eqref{vac} the winding number density $\nnw$ equals the Chern-Simons
number density $j^0_{\rm CS}$.

The winding number \eqref{nw} and Chern-Simons number \eqref{ncs} are
not gauge-invariant; they change by an integer under so-called large
gauge transformations. As a consequence a vacuum configuration can
have any integer winding number and Chern-Simons number, as long as
they are equal $N_{\rm CS}=N_{\rm w}$.  Under gauge transformations
$N_{\rm CS}$ and $N_{\rm w}$ change by the same amount, so that the
difference $N_{\rm w}-N_{\rm CS}$ is gauge-invariant.  The change in
time of $\NCS$ as defined by the integral over $q$ in
\eqref{intanomdiv} is also gauge invariant.

In the following we briefly discuss two well-known configurations that
can play a role in changing the Chern-Simons and/or the winding
number, namely the sphaleron and the texture.

\subsection{Sphaleron}
A sphaleron transition is a change from a vacuum with winding numbers
$N_{\rm CS}=N_{\rm w}=n$, to another vacuum with $N_{\rm CS}=N_{\rm
w}=n\pm 1$. It has been shown \cite{Manton:1983nd,Klinkhamer:1984di}
that the system must pass an energy barrier. The static and unstable
configuration at the minimum barrier height is called a sphaleron, and
its energy is the sphaleron energy.  This configuration has vanishing
Higgs length in the center, so that the winding number is not defined:
it jumps by an integer exactly at the transition.  The Chern-Simons
number of a sphaleron is precisely $1/2$ (up to an integer).

The sphaleron energy $E_{\rm sph}$ is proportional to $v/g$, and
approximately 10 TeV. Because of this high energy-barrier, tunneling
through the barrier (which corresponds to an instanton-like event) is
strongly suppressed.  Therefore the baryon number is effectively
conserved at low temperatures. At higher temperatures the suppression
is weaker because of thermal fluctuations over the barrier.  It is
also useful to interpret this in terms of an effective
temperature-dependent Higgs length $\langle\rh\rangle < v$ and an
effective barrier height $\propto \langle \rho\rangle /g$.  Above the
electroweak phase-transition temperature $\langle\rh\rangle$ vanishes
and sphaleron transitions occur unsuppressed.  During a tachyonic
electroweak transition there are also frequent fluctuations over the
barrier, as observed numerically in the susceptibility $\langle
\NCS^2(t)\rangle$ \cite{Garcia-Bellido:2003wd,Tranberg:2003gi}.

\subsection{Texture}
Without gauge fields, a texture is a configuration with a nonzero
winding number $N_{\rm w}$, with the Higgs length equal to the vacuum
value everywhere, and with only gradient energy.  According to
Derrick's theorem \cite{Derrick:1964ww} such a configuration is
unstable because its energy can be lowered indefinitely by shrinking
it.  Numerical simulations show that textures shrink quickly, and it
was argued in \cite{Turok:1989ai} that in the end the configuration
looses its winding number and decays into outgoing waves.

For the SU(2) Higgs model a natural extension of a texture is a gauged
texture: a configuration with Chern-Simons number different from the
winding number: $N_{\rm CS}-N_{\rm w} = \pm 1$.  One can think of an
initial configuration in which the gauge fields are pure-gauge with
integer Chern-Simons number and Higgs length equal to the vacuum
value.  Just as in the global case, a gauged texture is
unstable. There are basically two ways in which it can decay
\cite{Turok:1990in,Turok:1990zg}: when its size is smaller than
approximately $1/m_W$, it decays by changing the winding number, and
when it is larger it decays by changing the Chern-Simons number.  In
either case $N_{\rm CS}-N_{\rm w}\to 0$ and the configuration can
spread indefinitely into outgoing waves.

\subsection{Half-knot}
Although the total winding number in a finite volume with periodic
boundary conditions is integer, in practice there is no reason to find
local configurations with nearly integer winding number or
Chern-Simons number. This is because there is no mechanism that would
create such configurations, as there is, for example, energy
minimization for monopoles. Consequently the winding number density
can be spread out over the volume. However as we will argue below,
there will be high winding number density regions where the Higgs
length is very small. The total winding number in such a region is
typically not integer, but close to $1/2$, which is why we call these
configurations half-knots.

\paragraph{One dimension}
We illustrate this idea first in the simpler but analogous one
dimensional case with a complex scalar field $\Ph$ and global symmetry
group U(1),
\be
\Ph = \frac{1}{\sqrt{2}}\,(\ph_1 + i\ph_2) = \frac{\rh}{\sqrt{2}}\; \Om,
\quad
\Om \in U(1).
\ee
The winding number density is ($x\equiv x^1$)
\be
\nnw = -\frac{i}{2\pi}\, \Om^*\partial_x \Om
= \frac{1}{2\pi \rh^2}\, (\ph_1 \partial_x \ph_2 - \ph_2 \partial_x \ph_1).
\label{nnw1d}
\ee
In a coordinate patch were we can write
$\Om(x) = \exp[i\om(x) + \mbox{const.}]$
we also have 
$\nnw = \frac{1}{2\pi}\,\partial_x \om$.

In order to gain some intuition, let us consider the following
simple form
\begin{equation}
\phi_1(x) = \cos(x) - .95, \hspace{1cm} \phi_2(x) = \sin(x),  \label{1d}
\end{equation}
for which the Higgs length $\rho$ has a minimum when $x$ is close to
zero. This configuration is shown in a parametric plot in figure
\ref{fig:phis}. In this plot the Higgs length $\rho$ is the distance
from the origin, and the change of phase corresponds to the 'winding'
of the curve around the origin. When there is a small Higgs length,
the phase changes quickly (in this case approximately by an amount
$+\pi$) and there is a high winding number density. We can see this
also in figure \ref{fig:1dana}, where the Higgs length squared
$\rho^2$ and the winding number density are plotted. 
We call such a region with small Higgs length and large winding number
density a {\em half-knot}. Note that the total winding number is
integer (in this case $+1$), but that only part of the winding number
density is concentrated in a small region. The rest of is distributed
approximately homogeneously over the rest of space.

We can formalize the half-knot by approximating
$\phi_1$ and $\phi_2$ locally (around $x=0$) by a linear form
\be
\ph_\al = c_\al + d_\al x,
\quad
\al = 1,2.
\ee
This corresponds to approximating the circle near the origin by a
straight line, and gives
\bea
\nnw &=& \frac{1}{2\pi\rh^2}\, (c_1 d_2 - c_2 d_1),
\\
\rh^2 &=& c_\al c_\al + 2 c_\al d_\al x + d_\al d_\al x^2,
\eea
and a contribution to the winding number
\be
\nw^{\rm peak} \equiv \int_{-\infty}^{\infty} dx\, \nnw =
\half\, \mbox{sgn}(c_1 d_2 - c_2 d_1)
= \pm \half.
\ee
We have extended the integral to $\pm\infty$, but of course, the linear
approximation breaks down somewhere and the integral is to be interpreted as
the contribution from a peak in the winding density.

\paragraph{Three dimensions}
In this subsection we introduce half-knots for the three dimensional
case.  As in the one dimensional case we parametrize the Higgs field
by real fields 
\be 
\Phi = \frac{1}{\sqrt{2}}\, \left(\phi_4 1 + i
\phi_a \tau^a\right),
\ee
A simple example is 
a configuration that can locally be approximated by Fourier modes: 
\be
\phi_{\alpha} (x) = \sin(\vecx \cdot \veck_{\alpha} -
\epsilon_{\alpha}),
\quad \al = 1,\ldots,4.
\label{3d2}
\ee
The Higgs length $\sqrt{\phi_\al\phi_\al}$
will be small near the origin if all the
$\epsilon_{\alpha} \ll 1$. In order to get a local minimum and not a
long streamline of small Higgs length, the vectors $\veck_{\alpha}$ should
span three dimensional space.
In figure \ref{fig:3dana} we plotted the Higgs length and the winding
number density
as a function of $x=x^1$ and $y=x^2$,
in a slice through $z=x^3 = 0.05$,
for the case
\begin{gather}
\veck_1=(1,0,0),\;
\veck_2= (0,1,0),\;
\veck_3 = (0,0,1),\;
\veck_4=(0,0,1),
\quad
\ep_1=\ep_2=\ep_3=0.1,\; \ep_4=0 .
\label{3d4}
\end{gather}
The integrated winding
number in a box $x,y,z \in [-0.5,0.5]$ around the peak is found to
be $0.43$, and it does not depend strongly on the integration volume.

The 3d half-knot may be formalized similar to the 1D case
by using a linear approximation in a region where the Higgs length is small
(on the scale of $m_{\rm H}$),
\be
\ph_\al(\vecx) = c_\al + d_{\al k} x^k.
\ee
Then the winding number density is given by
\be
n_{\rm W} = \frac{1}{12\pi^2 \rh^4}\, \ep_{jkl}
\ep_{\al\bt\gm\dl}
\partial_j\ph_\al \partial_k\ph_{\bt}\partial_l\ph_\gm \,\ph_\dl,
\ee
and in the linear approximation this gives
\be
n_{\rm W} =\frac{1}{2\pi^2\rh^4}\,\det M,
\ee
where $M$ is the $4\times 4$ matrix consisting of the column
vectors
$d_{\al 1}$, $d_{\al 2}$, $d_{\al 3}$, $c_\al$,
\be
\det M =  (1/6) \ep_{jkl}\, \ep_{\al\bt\gm\dl}\, d_{\al j}\,
d_{\bt k}\, d_{\gm l}\, c_\dl.
\ee
The integral over the winding density can be done by
shifting coordinates, $\vecx\to \vecx'$,
\be
x^k = x^{\prime k} - g^{kl} d_{\al l}\, c_\al,
\ee
where $g^{kl}$ is the inverse of $f_{kl}$ defined by
\be
f_{kl} =  d_{\al k} d_{\al l},
\quad g^{kl} f_{lm} = \dl^k_m.
\ee
In terms of the shifted coordinates we have
\be
\ph_\al = c'_\al + d_{\al k} x^{\prime k},
\quad
c'_{\al} = c_\al - d_{\al k} g^{kl} d_{\bt l} c_\bt,
\quad
c'_\al d_{\al k} = 0,
\ee
and the length of the Higgs field is given by
\be
\rh^2 = c'_\al c'_\al
 + f_{kl}\, x^{\prime k} x^{\prime l}.
\ee
The winding number of the half-knot equals
\be
\intvecx n_{\rm W} = \frac{1}{2}\, {\rm sgn}\det M = \pm \half.
\ee
Since $f_{kl}$ is a positive matrix, the center
(maximum winding-number density) of the half-knot
is at $\vecx' = 0$, and it has an
ellipsoidal shape (surface of constant $\nnw$).
Its energy density has a constant contribution from the gradients,
$\half \partial_k\ph_\al\partial_k\ph_\al = \half d_{\al k} d_{\al k}$,
whereas the contribution from $\quart \lm(\rh^2-v^2)^2$ drops off
away from the center.

When $\rh$ vanishes in the center as a consequence of dynamics, so
when the vector $c_\al$ vanishes, the winding number may or may
not flip sign, depending on how the vector $c_\al$ recovers from
zero. A pure-Higgs half-knot can decay by spreading. With a gauge
field present, the Chern-Simons number may adjust to the winding
number locally, such that the difference between winding and
Chern-Simons number essentially vanishes.

Half-knots occur generically near the moment textures decay by
changing their winding number, or near sphaleron transitions, because
at these moments the Higgs length vanishes at a point. But half-knots
are more general, for example, they occur in random field
configurations, e.g.\ initial conditions for classical evolution.  It
is not clear yet at this stage that they are relevant, but in the
simulations we will see that they are.

\section{Winding in the tachyonic transition} \label{stt}
In this section we discuss the evolution of winding number and
Chern-Simons number in a fast tachyonic transition. First we will
review the relevant features of such a transition. After that we will
discuss the importance of half-knots and differentiate between early
and late half-knots.

\subsection{Tachyonic transition} \label{sec:tt}
At the onset of the tachyonic transition, when the effective mass
parameter $\mueff^2$ of the Higgs field changes sign, the universe is
assumed to be in a homogeneous state with $\langle\Ph\rangle\approx
0$. As $\mueff^2 \to -\mu^2$, the Higgs potential becomes unstable
near the origin and the low momentum modes of $\Ph$ grow very
fast. Since the couplings in the Standard Model are fairly weak, it
makes sense to study this process neglecting interactions. In this
approximation the Fourier modes of the Higgs field satisfy
\be
\ddot \Ph_\al(\veck,t) + [\mueff^2(t) + \veck^2]\Ph_\al(\veck,t) =
0,
\label{gaussappr}
\ee
which can be solved exactly for the initial stage where $\mueff^2
\approx -M^3 t$ \cite{Asaka:2001ez,Garcia-Bellido:2002aj} (choosing
$t=0$ as the onset of the transition). It turns out that the unstable
field modes, i.e.\ the modes with $\veck^2 < -\mueff^2$ grow very
fast; the number of unstable modes also grows when $-\mueff^2$
increases.

\paragraph{Interactions}
An estimate for the moment that interactions set in is given by the
time that the average Higgs field reaches the point where the second
derivative of the potential vanishes.  This is around $m_{\rm H}
t=4.8$, for an instantaneous quench and $m_{\rm H}/m_W = \sqrt{2}$
\cite{Tranberg:2003gi}. There are both self-interactions of the Higgs
field and interactions with the gauge field. The self-interactions
slow down the growth of the Higgs field, and eventually lead to an
oscillation near the vacuum state.  The interactions with the gauge
fields lead to a strong growth of the gauge fields
\cite{Skullerud:2003ki,Garcia-Bellido:2003wd}. The oscillation of the
Higgs field is damped by the interactions, and when more fields are
added in a realistic situation, this suppression is expected to be
even stronger.  Eventually the energy will be distributed over all
modes, and the system thermalizes.

\paragraph{Instantaneous quench}
As in \cite{Smit:2002yg,Skullerud:2003ki,Tranberg:2003gi},
we make in this paper the approximation
that the change of the potential is instantaneous in
order to obtain the most dramatic effects,
\bea
\mueff^2(t) &=& +\mu^2, t<0,\nonumber\\
&=& -\mu^2, t>0.
\eea
Moreover we do not consider the inflaton field in our simulations.
In this quenching approximation the modes of the Higgs field grow
exponentially fast, $\Ph_\al(\veck,t) \propto
\exp[\sqrt{\mu^2-\veck^2}\, t]$ \cite{Smit:2002yg,Tranberg:2003gi}.

\paragraph{Classical approximation}
Another approximation that we use is the classical approximation.
Intuitively one can see that the fields can be considered to be
classical, because the Bose fields grow exponentially fast and the
occupation numbers are therefore quickly much larger than one.  For
the gauge field the occupation numbers become substantial only after
the Higgs current in its equation of motion has grown sufficiently
large, which typically takes a few $m_{\rm H}^{-1}$ units of time
\cite{Skullerud:2003ki}.  The approximation is implemented as follows
\cite{Smit:2002yg,Tranberg:2003gi}.  Before the instantaneous quench
the fields are in the zero-temperature ground state corresponding to
positive $\mueff^2 = \mu^2$. Neglecting interactions this corresponds
to a gaussian distribution, which can be followed until it becomes
classical and a switch to classical evolution can be made.  However,
because quantum and classical evolution are formally the same for
gaussian systems, this switch can already be made at time zero,
directly after the quench.  The classical evolution is computed from
the fully non-linear equations of motion, including the
interactions. Making the switch early on also enables a more gradual
inclusion of the effect of the interactions.  We draw a number of 
Higgs field configurations from the classical part of the gaussian
distribution, and take these configurations as initial conditions for
the system after the quench.  For simplicity, the initial gauge
potentials are set to zero, whereas the $SU(2)$ electric fields are
calculated from Gauss' law \cite{Tranberg:2003gi}.  Then we evolve
each of these configurations according to the classical equations of
motion.  In the end we compute expectation values by averaging over
the initial distribution.

The classical approximation for a tachyonic transition has been
compared with quantum methods like the 2PI-method in
\cite{Arrizabalaga:2004iw}, and it turned out that
the two approximations agreed for the times and couplings used here,
giving further support for both.

\subsection{Winding and Chern-Simons number densities}
In the initial conditions for the tachyonic transition the gauge
fields are negligible, and since the gauge potentials are zero in our
implementation, the Chern-Simons number density is zero.  The Higgs
field initially has fluctuations around zero, and therefore it has
nonzero winding number density. Since the initial conditions are
random, the winding number density will be randomly distributed over
the volume. The total winding number in the volume will be integer,
and does not have to be zero.

When the system thermalizes and the temperature decreases, the
Chern-Simons number will approach the winding number.  If the winding
number would not change during the process, the Chern-Simons number,
approaching the initial winding number, would be determined by the
initial conditions, and CP violating interactions could not influence
the final outcome.

In reality the winding number does change during the process, and this
makes it possible that CP violation creates an asymmetry. The winding
number can change when the Higgs length becomes zero in a point, and
as we argued in the previous section there will be half-knots around
such points.  There are two periods when the Higgs has a chance to be
small and change of winding is likely to occur: early in the
transition when the Higgs field starts from a small fluctuation, and
later on, when the Higgs length bounces back due to its
self-interaction, or just any interactions, e.g.\ scattering of
non-linear waves.

In both periods half-knots will occur; we call them early and late
half-knots respectively.

\subsection{Early half-knots}
In the initial conditions of the tachyonic transition, the Higgs field
has small fluctuations around zero. The number density of minima of
the Higgs length is, depending on the initial conditions, roughly
proportional to $k_{\rm max}^3$ where $k_{\rm max}$ is the largest
wavenumber that is initialized.  Because of the peculiar feature of
the tachyonic transition that modes grow faster as their wavelengths
are larger, this number density of minima will quickly decrease. Hence
initially there are many half-knots, but their number quickly
decreases.

Some half-knots will manage to survive longer. When a half-knot still
exists when the gauge fields start to become important, the
Chern-Simons number density in these regions can adjust to the winding
number density.  When the Chern-Simons number becomes approximately
equal to the winding number in a blob, the covariant derivative $D_i
\Ph$ becomes small, the gradient energy diminishes and the half-knot
becomes stable.

The early half-knots are perhaps not so important for baryogenesis.
In principle CP violation could cause an imbalance in the formation
and decay of the number of half-knots and anti-half-knots. However in
this early period there are no interactions yet, and CP violation
cannot have acted. Also when the early half-knots stabilize and
survive CP violation is not important because then the winding number
does not change. So we expect for possible effects of CP violation, we
should look at the late half-knots.

\subsection{Late half-knots}
The Higgs length can also become small later in the transition. For
example this can happen when the Higgs field bounces back in its
potential, or because of interactions in general.  In this case there
will be late half-knots in which the winding can change. Because
interactions are important to create these half-knots, also
CP violating interactions can influence this process.  There may also
be longer lived half-knots, not stabilized by the gauge fields, whose
probability to decay is influenced by the stronger CP violation at
later times.

\section{Numerical simulations} \label{sns}
In this section we first describe briefly the setup of our
simulations, and then present the results.

\subsection{Setup}
In \cite{Tranberg:2003gi} numerical simulations were described with
the $SU(2)$ Higgs model, using the approximations described in section
\ref{sec:tt}, and with an extra CP violating term in the action. For
the present work we extended the computer code of
\cite{Tranberg:2003gi} to be able to observe the winding-number
density and a local quantity $\nncs$ closely related to the
Chern-Simons number density (see below).  We do not use the
CP violation of the code of \cite{Tranberg:2003gi} because at this
point we are interested in the mechanism of winding number and
Chern-Simons number production, and not yet in the creation of the
asymmetry.

The simulation was performed on a $60^3$ lattice, with periodic
boundary conditions and with a lattice spacing of 0.35~$m_{\rm
H}^{-1}$, such that the physical volume was $L^3 = (21\, m_{\rm
H}^{-1})^3$.  The initial conditions mentioned in \ref{sec:tt} are the
``just-a-half'' initial conditions as defined in
\cite{Smit:2002yg,Tranberg:2003gi}. Effectively this means that only
the growing modes, with momentum $k$ smaller than $\mu$, are
initialized with probability given by the vacuum state. Furthermore we
took $\lambda/g^2=1/4$, which is equivalent to $m_{\rm
H}/m_W=\sqrt{2}$.  (We shall also present some results for $m_{\rm
H}/m_W = 2$.)  For the determination of the initial conditions, which
are set by quantum fluctuations, we also have to fix $g^2$. We chose
$g^2=4/9$.  See \cite{Tranberg:2003gi} for more details on the
numerical implementation.

The density $\nncs$ is defined as
\be
\nncs(\vecx,t) = \int_0^t dt'\, q(\vecx,t'),
\ee
where $q$ is the gauge-invariant topological charge density given in
\eqref{qdef}. Since $q=\dmu j_{\rm CS}^\mu$ and the Chern-Simons current
is zero for our initial conditions,
\be
\nncs(\vecx,t) =
j^0_{\rm CS}(\vecx,t) + \partial_k \int_0^t dt'\, j^k_{\rm CS}(\vecx,t').
\ee
So $\nncs$ differs from $j^0_{\rm CS}$ by a divergence and
they both integrate to $\NCS$. In the following
we shall call $\nncs$ the Chern-Simons density, for simplicity,
but it should be
kept in mind that it is not equal to $j^0_{\rm CS}$.

\subsection{Results}

\subsubsection{One typical trajectory}
In order to investigate classical field configurations we look at
single trajectories. In this subsection we consider one typical
trajectory.

\paragraph{Total $N_{\rm w}$ and $N_{\rm CS}$ as function of time in
  typical run} The variables considered in \cite{Tranberg:2003gi} were
the
spatial average of the Higgs length squared
\be
\overline{\rho^2}= L^{-3} \intvecx \rh^2
,
\label{defhl}
\ee
the winding number $N_{\rm w}$ and the Chern-Simons number $N_{\rm
CS}$.  In figure \ref{fig:gen30} we show the evolution of these
variables in time for our typical trajectory, from $m_{\rm H} t =0$ up
to $m_{\rm H} t =30$. The winding number $N_{\rm w}$ fluctuates
initially, and later stays put at an integer.  The initial
fluctuations indicate that there must be zeroes in the Higgs length.
In the continuum these fluctuations would be between integers (a
`devil's staircase'), but here they appear as smoothed out by the
lattice discretization.

We also see that the Chern-Simons number starts only when the average Higgs
length is already rather large, and that at the later times
$N_{\rm CS} \approx N_{\rm w}$.
(Occasionally we also have seen trajectories for which the two
differed at $m_{\rm H} t =30$ by a number of order 1, and only at
much later times $\NCS$ approached $\nw$ (sometimes this took as
long as
$m_{\rm H} t \approx 500$).

\paragraph{3D pictures of $n_{\rm w}$ and $n_{\rm CS}$}
Next we look at the densities of the winding number and Chern-Simons
number in this trajectory. Figure \ref{fig:wind} displays the winding
number density in the three dimensional simulation volume from times
$m_{\rm H} t = 1$ to $m_{\rm H} t = 15$.  Note that the box has
periodic boundary conditions. Red (dark) indicates positive density,
blue (light) negative. In the beginning there are many 'blobs' in
winding number density. We will argue below for two specific cases
that these blobs are half-knots with a small Higgs length $\rho$ in
their center.  Sometimes they change sign. The number of blobs
decreases first until approximately time $m_{\rm H} t = 9$, then it
increases until approximately $m_{\rm H} t =13$ after which it
decreases again. Some of the early blobs that are there already from
the beginning survive all the time. The blobs that appear after time
$m_{\rm H} t = 9$ seem to be uncorrelated to the blobs that were there
before. We call these new blobs the late blobs.

In figure \ref{fig:ncs} the Chern-Simons number density is shown from
times $m_{\rm H} t = 7$ to $m_{\rm H} t =15$.  Before $m_{\rm H} t =7$
the Chern-Simons number density is negligibly small. Also in the
Chern-Simons number density there are blobs.  These blobs are
correlated with the winding number blobs.

\paragraph{Small Higgs length means a lot of winding}
We argued above that regions with small Higgs length have typically a
large winding number density. This is confirmed in the simulations.
In figure \ref{fig:nwphi} the absolute value of the winding number
density $|n_{\rm w}|$ is plotted versus the normalized Higgs length
$(\rho/v)^2$ for each point on the lattice. The configuration of the
typical trajectory at time $m_{\rm H} t = 6$ is used, when the gauge
fields are still unimportant. We see that $|n_{\rm w}|$ and
$(\rho/v)^2$ are correlated such that, when the Higgs length on a
lattice point is small, the winding number density is typically large.

A consequence of this correlation is that when the average Higgs
length is small, there will typically be more winding blobs. We saw
this already in the three dimensional pictures of the winding number
density: there were less winding blobs around time $m_{\rm H} t = 8$, when
the average Higgs length is large. We can show this more
quantitatively, by plotting 
\cite{Stephens:1999qv}
$\intvecx |\nnw|$
in figure \ref{fig:absnw}.
We see in this figure that the peak in
$\intvecx|n_{\rm w}|$ at $m_{\rm H} t \approx
20$ is much smaller than the corresponding one at $m_{\rm H} t\approx
12$. This agrees with the fact that there are much less lattice points
with small Higgs length at $m_{\rm H} t = 20$. We can also see this from the
histograms in figure \ref{fig:hist}. 
We call the blobs that are
created in minima of the Higgs length, second (third, \ldots) generation
blobs. 

\paragraph{Correlation between winding number and Chern-Simons
  number}
Because the Higgs and gauge fields interact, $n_{\rm w}$ and $n_{\rm
  CS}$ are correlated. This could already be seen in the three
dimensional pictures, but we can also calculate the correlation
\be
C(\vecr,t) =
\frac{\intvecx[\nncs(\vecx,t)-\overline{\nncs}(t)]
[\nnw(\vecx+\vecr,t)-\overline{\nnw}(t)]}
{\sqrt{\intvecx[\nncs(\vecx,t)-\overline{\nncs}(t)]^2}
\sqrt{\intvecy[\nnw(\vecy,t)-\overline{\nnw}(t)]^2}},
\label{wcscorrvsr}
\ee
where the `over-bar' denotes the spatial average, as in \eqref{defhl}.
This correlator is plotted versus $r=|\vecr|$ at various times in figure
\ref{fig:wcscorrvsr}. It shows a spatial correlation developing
on distances of order $m_{\rm H}^{-1}$, modulated in time and showing
a tendency to diminish at later times.

It is also instructive to plot its value at $r=0$ versus time, see
figure \ref{fig:wcscorr}.  We see that the correlation
$C({\bf 0},t)$ develops already at early times, it peaks at times
$m_{\rm H} t \approx 12$ and 16, and there is a rapid drop after the
first peak. This drop occurs when the average Higgs length has become
small after its first maximum, and $\intvecx |n_{\rm w}|$ is on the
rise again (cf.\ figure \ref{fig:nwphi}). We interpret this as being
caused by the creation of many new winding blobs when the Higgs length
is small again, half-knots that are uncorrelated with the $\nncs$.
When $C({\bf 0},t)$ peaks for a second time the average Higgs
length is large again and $\intvecx |n_{\rm w}|$ is low.  We suspect
that this is because the winding blobs that still exist when the
average Higgs length is large, exist already for some time and the
Chern-Simons number density has had some time to adjust.  Later on the
correlation decreases, which is presumably caused by random
fluctuations.

In the following two subsections we zoom in on two blobs, first on an
early blob and then on a late survivor.

\subsubsection{Early blob}
For the early blob we take the one indicated by the arrow in figure
\ref{fig:blob2}.  Let us first look at the distributions of the Higgs
length and the winding number density in this blob. In a vertical
slice in the $xz$-directions through the center of the blob, the Higgs
length and the winding number density are plotted in figure
\ref{fig:slice2}, at time $m_{\rm H} t =2$.  The Higgs length has a
minimum and the winding number has a large peak at this minimum.
These figures look very similar to the analytical example in figure
\ref{fig:3dana}.

Next we have calculated the sums of some quantities in a ball around
the center of the blob. For this we had to determine the position of
the center, which is slightly ambiguous. We did it by defining the
center as the point where the winding number density is maximal. The
position of the center can change a bit at different times, so we
determined the center at each time step.  In the left panel of figure
\ref{fig:ball} we show the integrated winding number density
\be
\nw^{\rm ball} = \int_{\rm ball} d^3 x\, \nnw,
\ee
integrated Chern-Simons number density
\be
\NCS^{\rm ball} = \int_{\rm ball}d^3 x\,\nncs,
\ee
and the volume-averaged Higgs length
\be
\overline{\rh^2}^{\rm ball} =
\int_{\rm ball}d^3 x\, \rh^2
\Big/
\int_{\rm ball}d^3 x\, 1
,
\ee
for a ball of radius 6 lattice units, corresponding to 2.1~$m_{\rm
H}^{-1}$, as a function of time.  For reference the Higgs length
averaged over the full simulation volume is also shown.

The winding number in the ball first decreases until $m_{\rm H} t =
5$, then increases until $m_{\rm H} t = 10$ and afterwards stays
approximately constant near a value of 0.3. The Chern-Simons number
becomes visible from $m_{\rm H} t \approx 8$ onwards, has a small peak
and stays constant near 0.2 after $m_{\rm H} t \approx 13$.  The
winding number and Chern-Simons number end up being close to each
other. The average Higgs length in the ball grows only much later than
the one in the full volume, and also oscillates with a somewhat higher
frequency.  It also exhibits much less damping, which is suggestive of
oscillons \cite{Broadhead:2005hn,Farhi:2005rz}.\footnote{The Higgs
mode of the ideal oscillon found \cite{Farhi:2005rz} for $\mh = 2\,
m_W$ oscillates at a slightly lower frequency than $\mh/2\pi$ but in
our case the effective Higgs mass will be lowered by a non-zero
effective temperature in the bulk.}  We will comment later on the dip
in the winding number at time $m_{\rm H} t =5$.

The right panel of figure \ref{fig:ball} shows the energy in the same
ball with radius 2.1~$m_{\rm H}^{-1}$.  We display the excess energy
above the average energy relative to the sphaleron energy, i.e.\
$\int_{\rm ball}d^3 x\,(\ep-\overline{\ep})/E_{\rm sph}$, where $\ep$
is the energy density and $\overline{\ep}$ its average over the total
volume. The average energy density is simply that of the origin of the
Higgs potential, $\overline{\ep}=\mh^4/16\lm$, and the sphaleron
energy for $\mh=\sqrt{2}\, m_W$ is $E_{\rm sp}\approx 3.78\, (4\pi
m_W/g^2)$ (see e.g.\ \cite{Baacke:1993aj}), and so $\int_{\rm ball}
d^3 x\, \overline{\ep}/E_{\rm sp} \approx 0.29$.  Hence, the sphaleron
energy in this plot is at 0.71.

We show the total energy in the ball as well as its contributions from
the Higgs and the gauge fields (the contribution from the covariant
derivative is allocated to the Higgs fields). We see that the gauge
fields contribute most to the energy. It is remarkable that the peak
in the total energy occurs at a time where the average Higgs length in
the ball has its first maximum, that the peak is significantly higher
than the sphaleron energy, and that the energy has already fallen back
to the average already shortly after $m_{\rm H}$t = 15. Evidently, a
strong energy flow into and out of the ball is taking place. At the
later times $\NCS^{\rm ball}$ has roughly the same value as $\nw^{\rm
ball}$.

To see how these results depend on the radius of the ball we show in
figure \ref{fig:varrad} the winding number and Chern-Simons number in
balls with increasing radii, from 3 up to 15 lattice units.  They
clearly depend on the radius, and for the larger balls $\nw^{\rm
ball}$ increases above 1/2. This may be caused by another blob of the
same sign that is close (cf.\ figure \ref{fig:wind}, e.g.\ at time
$\mh t = 15$ the distance between the centers of the two blobs is
about 13 lattice units).

Here we return to the dip that we observed at $m_{\rm H} t=5$ in
figure \ref{fig:ball}. From figure \ref{fig:varrad} we see that the
dip is also there for larger radii of the ball; apparently the winding
number is not flowing out of the ball, but is really decreasing. In
the continuum this can only occur when the Higgs length $\rho$ is
exactly zero somewhere. But on the lattice the we will miss already a
significant amount of winding number when the spatial size of the
winding number peak becomes smaller than a lattice unit. Hence we
interpret the observed dip as a lattice artefact, signaling a
half-knot in the center of which the Higgs length decreases (which
makes the peak sharper) until $m_{\rm H} t=5$, and increases again
after that.

Further insight can be obtained from the profiles of the Higgs
length and the winding-number density around the center of the
blob, $\rh(r)$ and $\nnw(r)$. They are plotted in figures
\ref{fig:1hp} and \ref{fig:wp},
for times $m_{\rm H} t =1$ to 10.
The profiles are determined by averaging at fixed distances $r$
from the center over all directions. For the position of the
center we used the same values as in figure \ref{fig:ball}.

From the Higgs length profiles we see that, while the Higgs length in
the bulk grows steadily from the beginning, in the center of the blob
it remains very small up to time $m_{\rm H} t\approx 6$, and starts to
grow only after that. At the latest time $\rh(r)$ looks like an
oscillation about the equilibrium value $\approx v$. 
The winding number density profile
is already well defined at time $\mh t = 1$. It then shrinks and
becomes steeper towards the center, $\mh t = 2$ and 3. This
shrinking and steepening appears to get blurred by lattice
artefacts at
$\mh t = 4,5,6$ (note that the profiles here are shown on a much smaller
scale than the $\rh$-profile in figure \ref{fig:1hp}), and as
mentioned earlier, we believe this is the reason for the dip in
$\nw^{\rm ball}$ at $\mh t = 5$.
From time $m_{\rm H} t=10$ the winding number
profile
broadens.

We conclude that we have witnessed the formation of a half-knot, that
nearly decayed by shrinking, but got `saved' by the gauge field
adjusting its Chern-Simons number density and diminishing the Higgs
gradient density $|D_i \Ph|^2$.  Remarkably, this adjustment goes
together with a big jump in gauge-field energy. At later times the
well-dressed blob carries no excess energy, and the process has led to
a local change in the total Chern-Simons number.

\subsubsection{Late transition}

Above we have seen (in the three dimensional pictures \ref{fig:wind},
\ref{fig:ncs} and the $|n_{\rm w}|$ graph \ref{fig:absnw}) that new
blobs are created when the average Higgs length is small. Sometimes
the winding number $N_{\rm w}$ changes in such a blob. Here we present
an example of such a late blob in which the winding number changes. It
comes from another trajectory than the one used before.  Figure
\ref{fig:gen31} shows the evolution of $\overline{\rh^2}$, $\nw$ and
$\NCS$ in this run up to time $m_{\rm H} t =30$.  Note that the
winding number changes from $-1$ to 0 between $m_{\rm H} t =23$ and
$m_{\rm H} t =24$.  This change takes place in the blob that we are
going to consider.

Figure \ref{fig:3drun31} shows 3D plots of the winding and
Chern-Simons number densities at times $m_{\rm H} t =23$ and 24.  The
change of the winding number occurs in the blob that changes sign at
the top of the box. At the same position there is a positive
Chern-Simons number density both before and after the change of
winding number.

In figure \ref{fig:nwdenslb} the Higgs length and the winding number
density in a horizontal slice through the blob are shown for times
$m_{\rm H} t =23$ and $m_{\rm H} t =24$.  There is a pronounced
minimum in the Higgs length at the place of the peak in the winding
number density. The latter changes quite abruptly from negative to
positive.

Next we plot $\nw^{\rm ball}$, $\NCS^{\rm ball}$ and
$\overline{\rh^2}^{\rm ball}$ for a ball of radius 6 in lattice units
($2.1 m_{\rm H}^{-1}$) as a function time in the left panel of figure
\ref{fig:balll}. The average Higgs length in the ball is approximately
in anti-phase with the average Higgs length in the full volume, and
there appears to be no damping, suggesting as in figure \ref{fig:ball}
a connection with the oscillon phenomenon
\cite{Broadhead:2005hn,Farhi:2005rz}. The winding number flips sign
around $\mh t = 2$ and becomes negative. Then it makes limited
excursions, even at the times where there are large peaks in
$\overline{\rh^2}^{\rm ball}$, but between times $m_{\rm H} t =23$ and
$m_{\rm H} t =24$ it makes a rapid jump by about $+0.6$, a substantial
part of 1 for this relatively small ball. At this point
$\overline{\rh^2}^{\rm ball}$ has a minimum. The Chern-Simons number
of the ball does not follow the winding number very much.  It shows
mild negative peaks at $\mh t = 9$ and 18, shortly before the peaks in
$\overline{\rh^2}^{\rm ball}$, and between $\mh t = 18$ and 30 it
gradually increases by about 0.6 (about the same as the jump in
$\nw^{\rm ball}$ at $\mh t = 23$.  In the right panel of figure
\ref{fig:balll} the total energy in the ball and the contributions
from the gauge and the Higgs fields are plotted versus time. As in
figure \ref{fig:ball}, we display the excess energy above the average,
and with respect to the sphaleron energy. The contribution from the
gauge fields is again dominant in the first two peaks (which coincide
with the peaks in $\overline{\rh^2}^{\rm ball}$), but at $\mh t = 23$
(where $\overline{\rh^2}^{\rm ball}$ has a minimum) the Higgs energy
clearly dominates.  There is a moderate rise of the energy between
$\mh t =22$ and 27.  Given that the subtracted energy is about 0.29
$E_{\rm sp}$, its maximum value is about 15\% higher than the
sphaleron energy.  The $\nw^{\rm ball}$ and $\NCS^{\rm ball}$ data for
balls with increasing radii are given in figure \ref{fig:varradl}. The
result is comparable to the early blob: both the winding number and
the Chern-Simons number increase with increasing radius, indicating
that there is not a sharp boundary of the blob.  However, the sharp
rise in $\nw^{\rm ball}$ between $\mh t = 23$ and 24, and the steady
increase of $\NCS^{\rm ball}$ after $\mh t = 18$, are present for all
ball radii.

Figures \ref{fig:latehp} and \ref{fig:latewp} show the profiles of the
Higgs length and the winding number density, from times $m_{\rm H} t
=19$ to $m_{\rm H} t=27$.  The Higgs length at the center is
decreasing and apparently developing a zero at time $m_{\rm H} t =23$,
when the winding number changes, and after that it increases
again. The winding profile becomes very steep around this time, as we
saw also in figure \ref{fig:nwdenslb}.  Lattice artefacts do not seem
to be prominent in this case.  Afterwards the winding density spreads
and becomes very small.

The transition at $\mh t=23$ bears the hallmarks of a sphaleron
transition: a gradual $O(1)$ increase in $\NCS$ and an $O(1)$ jump in
$\nw$, which occur locally in a blob, in the center of which $\rh$
goes through zero, together with a gradual increase in $\NCS^{\rm
bal}$ and a switch of sign in $\nw^{\rm ball}$. The energy at that
time in the ball of radius $2.1\,\mh^{-1}$ ($1.5\, m_W^{-1}$) is also
reasonably close to the sphaleron value ($\approx 0.9\, E_{\rm sp}$).
The properties of the subsequent maximum at $\mh t = 27$ look rather
similar to the two earlier ones, in its dominance of the gauge-field
energy and the accompanying maxima in $\overline{\rh^2}^{\rm ball}$.

\subsection{Distributions and susceptibilities}
Here we present some quantitative results for the late
distribution of winding numbers, and for the growth of the
Chern-Simons susceptibility
$\langle\NCS^2\rangle$ during the transition.
The winding-number distribution is expected to be gaussian for
large volumes, but its volume dependence may contain non-trivial
deviations. The rate of change of the Chern-Simons susceptibility
has been interpreted as an effective sphaleron rate and used
\cite{Garcia-Bellido:2003wd} to estimate the asymmetry induced by
CP violation.
In this section we also show results for mass ratio
$\mh = 2\, m_W$, in addition to the value $\mh = \sqrt{2}\, m_W$
used throughout this article. We vary $\mh/m_W$ by varying the
Higgs self coupling $\lm$ while keeping fixed the gauge coupling
$g^2$, the volume in Higgs mass units, $(\mh L)^3$, and the
lattice spacing in Higgs mass units, $a\mh$.

\subsubsection{Winding distribution}
Figure \ref{fig:distnw} shows the normalized distribution of
winding numbers at $\mh t=50$ obtained from a sample of about 2000
initial conditions for each parameter set. Four fits to the 
data are shown as well, one based on a
gaussian and three models based on a generation via winding blobs.
In a first approximation we treat such blobs as being dilute and
independent, which means that in a sufficiently large volume the
probability for $n$ blobs is $p_n = c r^n/n!$, where $c$ is a
normalization constant following from $\sum_n p_n = 1$. For
$n=1,2,\ldots$ this gives
$c=e^{-r}$, and $r$ is the average number of blobs, which is
proportional to the volume.

From a Kibble mechanism viewpoint
one might expect each blob
to contribute one unit to the winding number. For
$n$ blobs there may be $k$ blobs contributing +1 and $n-k$ blobs
contributing $-1$, such that the winding number is $N = k(+1)
+(n-k)(-1) = 2k-n$. Assuming a  probability $a$ for +1 and $(1-a)$
for $-1$, the probability for winding number $N$ would be given by
\be
P_N = \sum_{n=0}^{\infty} p_n\sum_{k=0}^n
\left(\begin{array}{c}n\\k\end{array}\right)
\dl_{2k-n,N}\,a^k(1-a)^{n-k}= e^{-r}
\left(\frac{a}{1-a}\right)^{N/2} I_N(2r\sqrt{a(1-a)}),
\ee
where $I_N$ is the usual Bessel function. In our case of no CP
violation, $a=1/2$, and
\be
P^{(1)}_N(r) = e^{-r} I_N (r).
\ee
For $r\gg 1$ this becomes indistinguishable from a gaussian,
\be
P^{\rm gauss}_N(\sg) =\frac{1}{\sqrt{2\pi \sg^2}} \,
e^{-N^2/2\sg^2},
\ee
with $\sg\approx r$.

However, we have argued and presented evidence that in a tachyonic
quench the initial winding blobs are half-knots, some of which
become stabilized by the gauge field and pick up a Chern-Simons
number equal to their winding number $\pm 1/2$. So their initial
winding number is conserved, although they later decay by
spreading. This suggest that we modify the above model by taking
into account the half integer winding of the blobs. Since the
total winding number is integer, we could modify the above
reasoning by assuming that in case $n$ is odd, there is a
compensating contribution $\pm 1/2$ somewhere in the volume,
writing
$N = k(+ 1/2) + (n-k)(-1/2) \pm 1/2$, with equal probability 1/2 for
the $\pm$ sign. The even-$n$ contribution to $P_N$ is unmodified.
This gives
\be
P^{(1/2)}_N(r) = e^{-r}\left[I_{2N}(r) + \half I_{2N+1}(r) +
\half I_{2N-1}(r)\right]. \label{hk1}
\ee
Alternatively, we can model the compensating $\pm 1/2$
contribution by a half-knot and only allow even $n$, such that
$p_n\to r^n/n!\cosh r$, which leads to the simpler form
\be
P^{\prime(1/2)}_N(r) = I_{2N}(r)/\cosh(r). \label{hk2}
\ee
The distributions $P_N$ are normalized, $\sum_{N=-\infty}^{\infty}
P_N = 1$. 

The $\ch^2$ values of the fit presented in the upper plot of
figure \ref{fig:distnw} for the half-knot based model of equation
\eqref{hk2} is clearly lower than the integer model and also the
gaussian model. For the model of equation \eqref{hk1} the result is
comparable. For the lower plot the integer-knot model gives a better
fit but the difference with the half-knot model is not significant
($\ch^2/\; {\rm d.o.f.}=1.3 \; {\rm vs.} \; 1.1$). This we consider
additional support for the relevance of half-knots in the tachyonic
transition.

\subsubsection{Chern-Simons susceptibility}

Figure \ref{fig:suscept} shows the time dependence of
$\langle\NCS^2\rangle$ for the mass ratios $\mh/m_W = \sqrt{2}$
and 2. Both curves show an initial rapid rise, and the $\sqrt{2}$
case shows a deep dip near $\mh t = 13$. This is about the time
where the average Higgs length also has its first minimum
(actually approximately one $\mh^{-1}$ unit later) . At this time
$\intvecx |\nnw|$ has risen again substantially (figure
\ref{fig:absnw}), and there is evidently no instantaneous
connection with the winding number. The dip is much less
pronounced (and shifted) for mass-ratio 2 case, presumably due to the
stronger coupling
$\lm$, which implies a smaller initial energy density
($\mh^4/16\lm$).
We have seen that the dip in the average Higgs length is also less
deep in this case. This suggests fewer second-generation winding
blobs, which may explain the quicker flattening of
$\langle\NCS^2\rangle$, compared to the $\sqrt{2}$
case. Correspondingly, the effective sphaleron rate
$d\langle\NCS^2\rangle/dt$ (e.g.\ averaged over an oscillation) will
be substantial over a larger time span when $\mh/m_W$ decreases.

An alternative interpretation for the first minimum in the
susceptibility could be given in terms of $\NCS$ bouncing back
from a barrier in the potential of its effective equation of
motion. The $\rh$-dependence of this barrier may even lead to
resonant behavior \cite{Smit:2002yg,Tranberg:2003gi}.

\section{Summary and discussion} \label{sd}

In the theory of baryogenesis the change of the Chern-Simons
number of the $SU(2)$ gauge field plays an important role, and we
studied the mechanism by which this can occur in a tachyonic
electroweak transition. The tachyonic instability occurs initially
in the Higgs field, and because of its coupling to the gauge field
through the covariant derivative, one expects a correlation
between the Chern-Simons number and the Higgs winding number. We
argued that in a tachyonic transition there will be many places
where the Higgs length is small in a typical field configuration.
These places are important, since the winding number can change
when the Higgs length goes through zero, possibly under influence
of CP violation, and this may also induce a change in the
Chern-Simons number. On the other hand, small Higgs lengths imply
small Higgs currents, which may limit their influence in the
equation of motion of the gauge field. Regions with small Higgs
length have in general a large winding-number density, which is
why we call them winding blobs.

The integrated winding number in these blobs does not need to be
integer, and the basic objects have winding number 
close to
$\pm 1/2$, the half-knots. When the dynamics causes the
Higgs length to vanish in the center of a half-knot, its winding
number may flip sign. Half-knot configurations occur also
naturally during sphaleron transitions, and decaying textures
loosing their winding number, since these have a moment at which
the Higgs length vanishes at a point in space.
The pure-Higgs half-knots can evaporate by increasing the Higgs
length in the center, but they may also get `dressed' by the gauge
field adjusting its Chern-Simons number density locally to the
winding. The configuration may then decay by spreading into the
environmental fluctuations, and the half-knots have acted like
local seeds of Chern-Simons number change.

We observed the winding blobs in numerical simulations of the
tachyonic transition in the SU(2) Higgs model. Because of their
large winding number density, they are easy to spot. We indeed
observed a strong correlation between the half-knot winding
density with the Chern-Simons number density.\footnote{We recall
that in our numerical simulation we actually used $\nncs$, which
is a gauge-invariant modification of
$j_{\rm CS}^0$ with the same total Chern-Simons number.}
The picture sketched above was supported by the behavior of the
integrated winding and Chern-Simons densities in small balls, as
well as the radial profiles of the spherically averaged densities. Our
findings for the profiles are similar to the one shown in
\cite{Garcia-Bellido:2003wd}. 

We also analyzed an example of a realistic sphaleron transition.
This occurred quite late in a blob that survived a relatively long
time, showing signs of stability reminiscent of oscillons
\cite{Broadhead:2005hn,Farhi:2005rz}. In the present case
we do not expect such objects to live very long as they will be
destroyed by thermal fluctuations.

We found that the winding blobs can be divided into two classes.
The early blobs are remnants from the initial conditions, and can
sometimes survive when they are stabilized by the gauge fields.
The late blobs occur when the Higgs length bounces back to small
values, and there can be second, third,
\ldots, generations, especially for smaller Higgs
self-couplings. Most of the early blobs are probably not important
for CP violation, because interactions become important too late
for them. CP violation can 
however affect the late blobs.

The distribution of Chern-Simons numbers is expected to be
approached by the distribution of winding numbers when the volume
becomes large. We studied the winding-number distribution and
found that it could be fitted by models based on half-knots,
better than by a model based on integer components and even
marginally better than a gaussian, although for large volumes all
the model-distributions are expected to become indistinguishable
from a gaussian.

Finally we presented new results on the susceptibility of
Chern-Simons numbers, which has been used in estimates of the
baryon asymmetry \cite{Garcia-Bellido:2003wd}. Some aspects of its
dependence on the Higgs self-coupling could be interpreted in
terms of generations of half-knots, but a detailed understanding
is difficult.
Nevertheless, we expect that the increased understanding obtained
in this paper is of use for modeling cold electroweak
baryogenesis.

\acknowledgments 
We thank Alejandro Arrizabalaga for useful suggestions. This work
received support from FOM/NWO. DS was supported by the ERASMUS
scholarship. AT is supported by PPARC SPG {\it``Classical lattice
field theory''}.

\bibliography{refs}

\newpage

\FIGURE{
\includegraphics[width=0.56\textwidth]{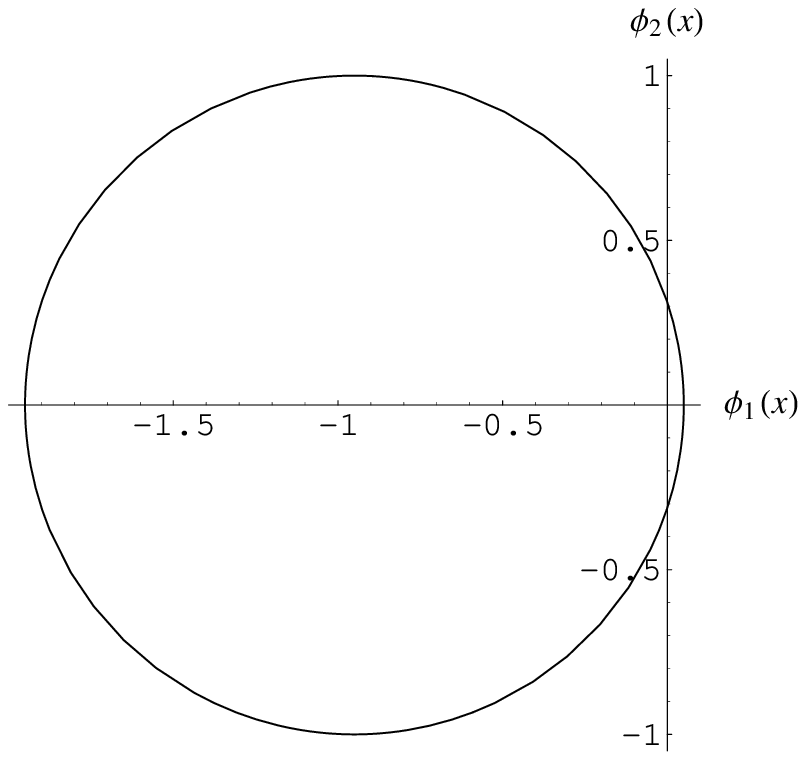}
\caption{Parametric plot of
$\phi_2(x)$ versus $\phi_1(x)$
for $x \in [-\pi,\pi]$, for the configuration of equation
  \ref{1d}. 
}
\label{fig:phis}
}

\FIGURE{
\includegraphics[width=.49\textwidth]{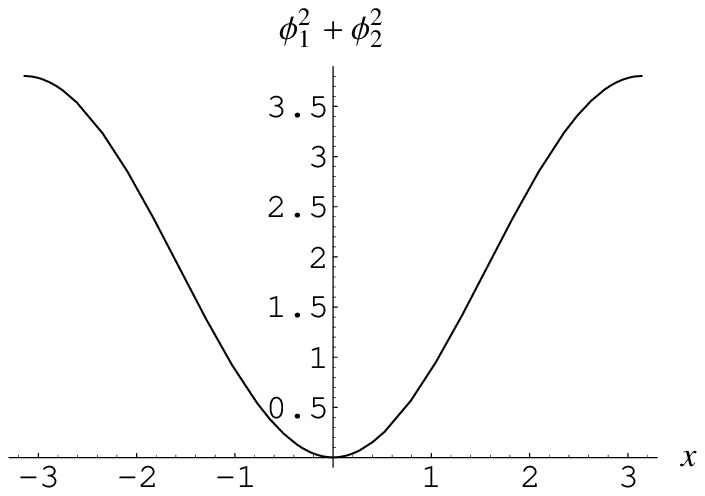}
\includegraphics[width=.49\textwidth]{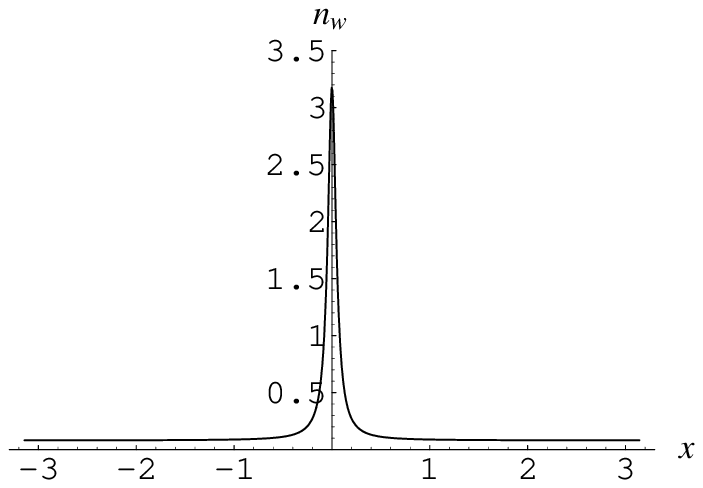}
\caption{The Higgs length $\rho^2$ and the winding number density
  $n_{\rm w}$ for the
  configuration of equation \eqref{1d} as function of
  $x$. 
}
\label{fig:1dana}
}

\FIGURE{
\includegraphics[width=.49\textwidth]{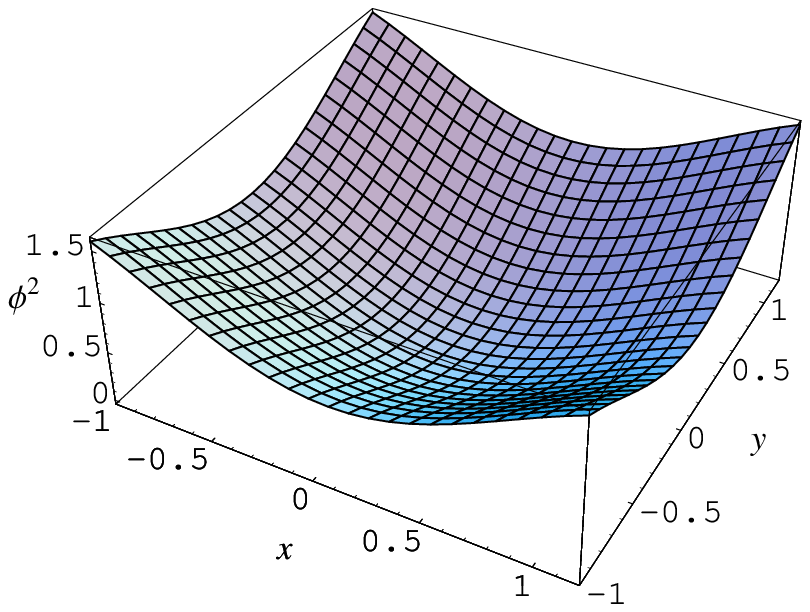}
\includegraphics[width=.49\textwidth]{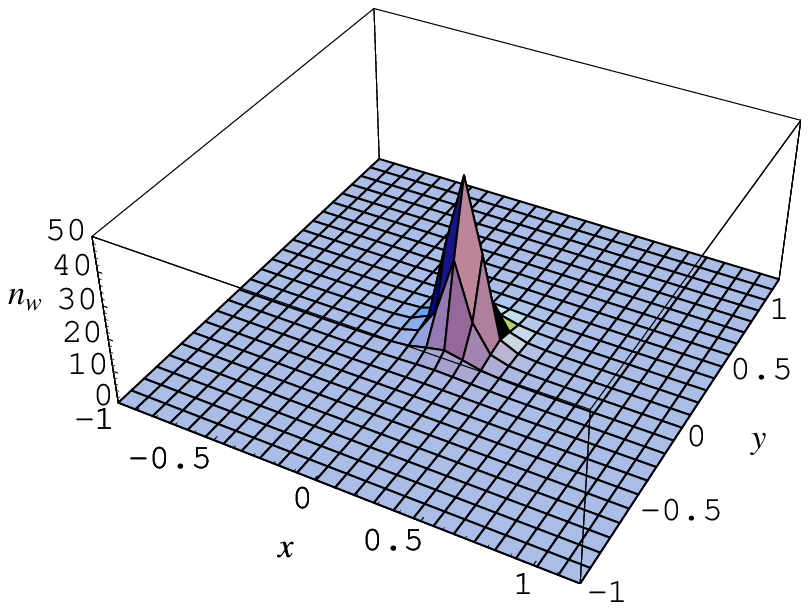}
\caption{The Higgs length $\rho^2$ (left) and the winding number
  density $n_{\rm w}$
  (right) for the configuration
\eqref{3d2},
\eqref{3d4},
as function of $x$ and $y$, with $z=0.05$.}
\label{fig:3dana}
}

\FIGURE{
\includegraphics[width=0.74\textwidth]{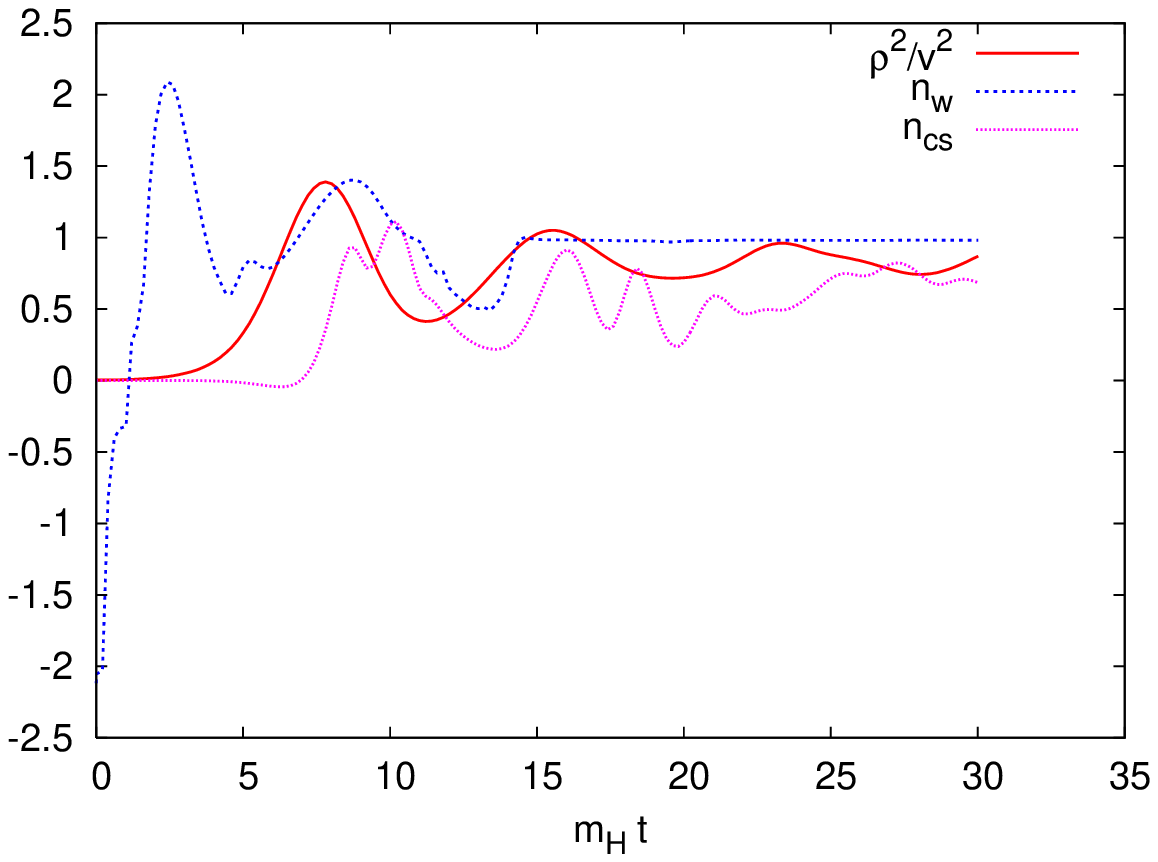}
\caption{Results of a typical run for times
   $m_{\rm H} t =0$ to $m_{\rm H} t =30$.
   Plotted are $\overline{\rh^2}/v^2$,
   the total winding number
   $N_{\rm w}$ and total Chern-Simons number $N_{\rm CS}$.}
\label{fig:gen30}
}

\FIGURE{
\includegraphics[width=10cm]{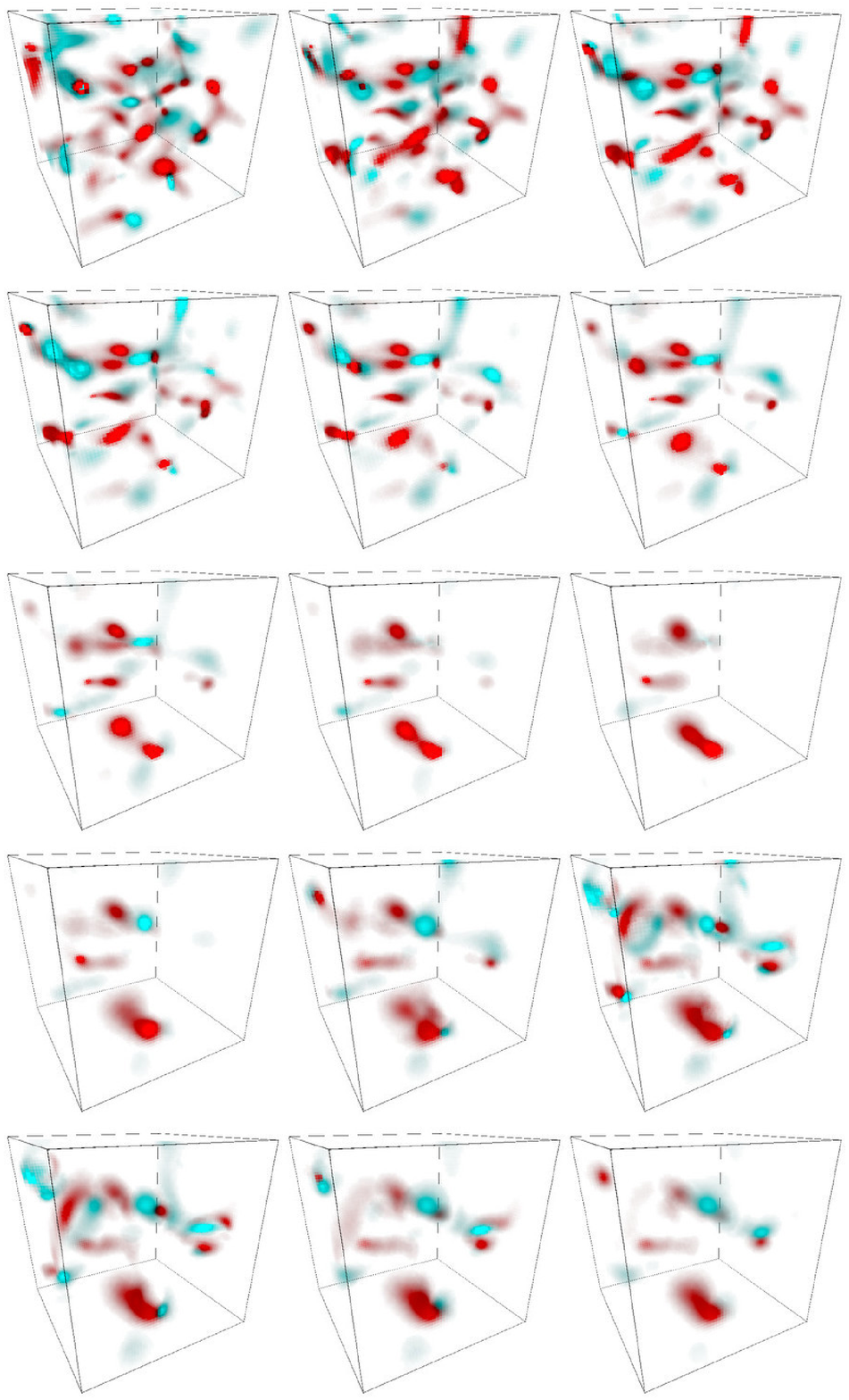}
\caption{Winding number density $n_{\rm w}$ from $m_{\rm H} t=1$ to $m_{\rm H}
  t=15$. Red (dark) is positive, blue (light) is negative.}
\label{fig:wind}
}

\FIGURE{
\includegraphics[width=10cm]{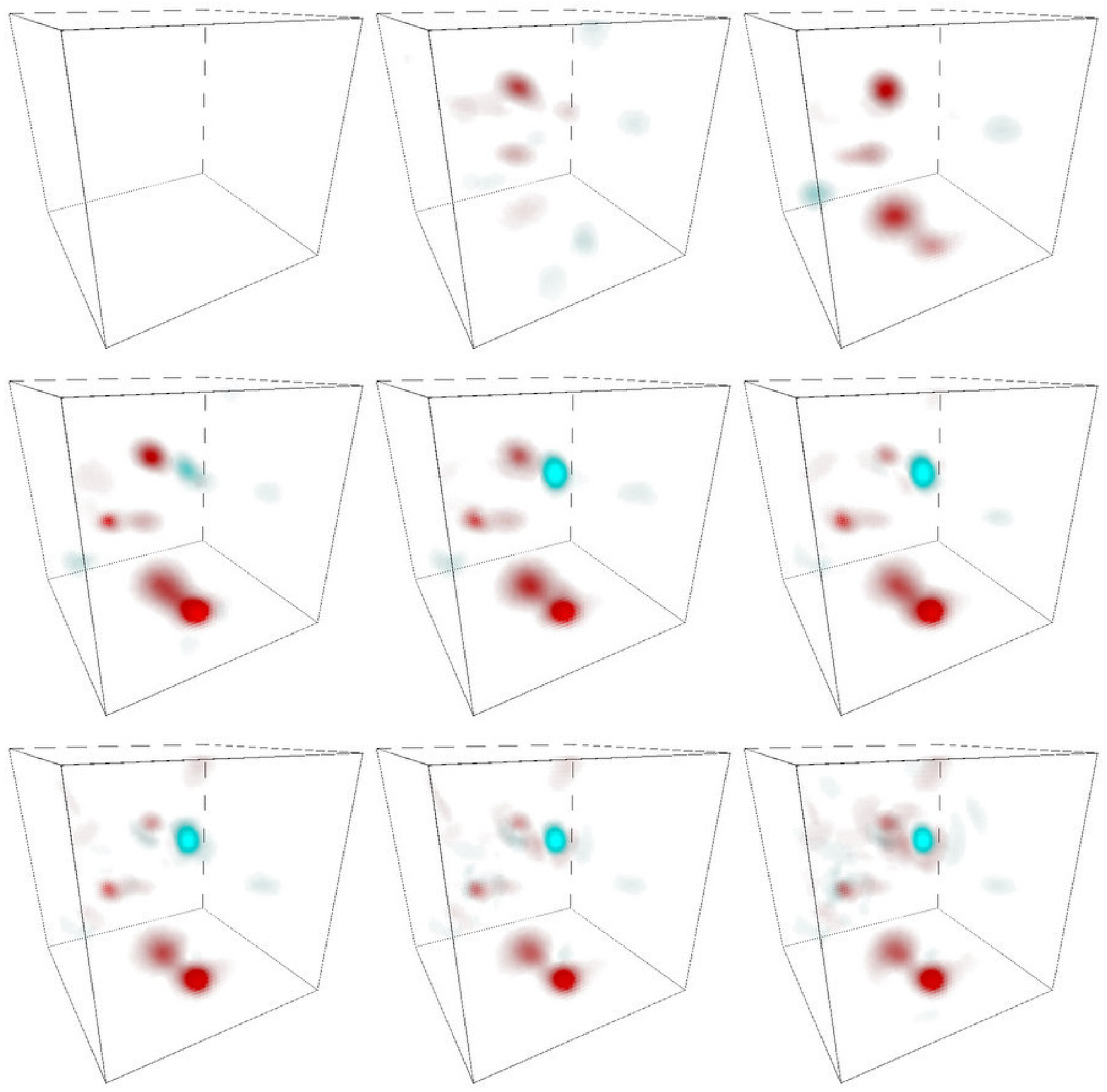}
\caption{Chern-Simons number density $n_{\rm CS}$ from
  $m_{\rm H} t=7$ to $m_{\rm H} t=15$. Before $m_{\rm H} t=7$ the Chern-Simons number
  density is negligibly small. Red (dark) is positive, blue (light) is
  negative.}
\label{fig:ncs}
}

\FIGURE{
\includegraphics[width=0.74\textwidth]{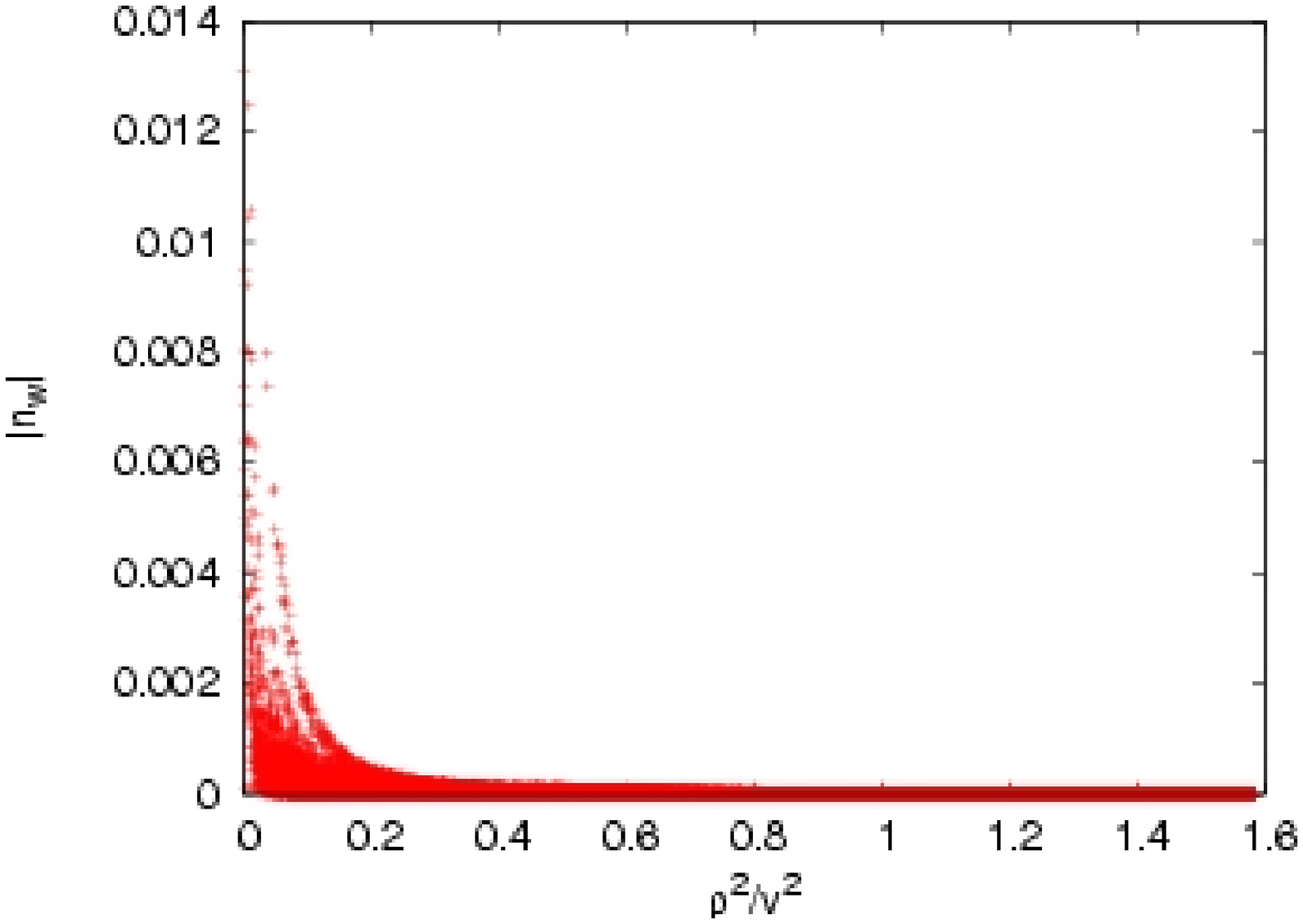}
\caption{The absolute value of the winding number density $|n_{\rm
    w}|$ versus the Higgs length for all lattice points in the
    simulation volume,
    at time $\mh t = 6$.}
\label{fig:nwphi}
}

\FIGURE{
\includegraphics[width=0.745\textwidth]{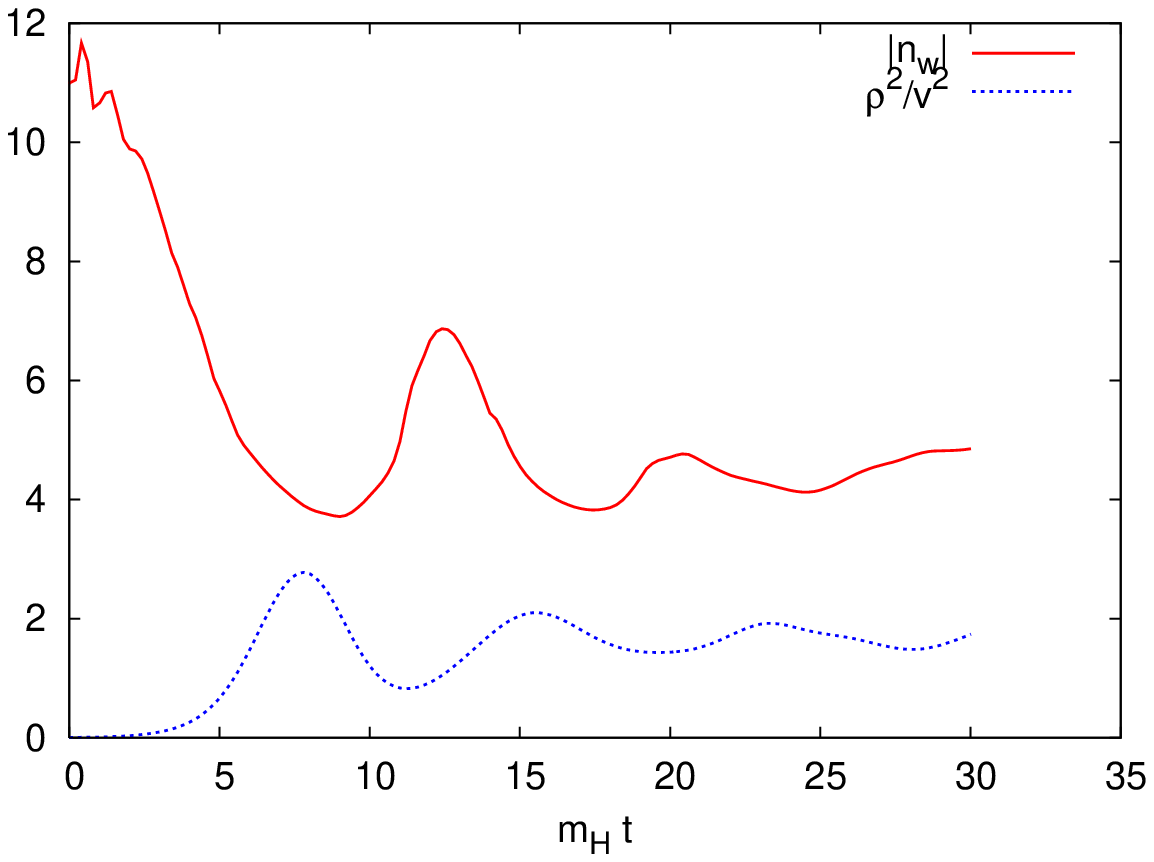}
\caption{
The integral $\intvecx |\nnw|$ and the spatial
    average of the squared Higgs length versus $m_{\rm H} t$.
    }
\label{fig:absnw}
}

\FIGURE{
\includegraphics[width=0.74\textwidth]{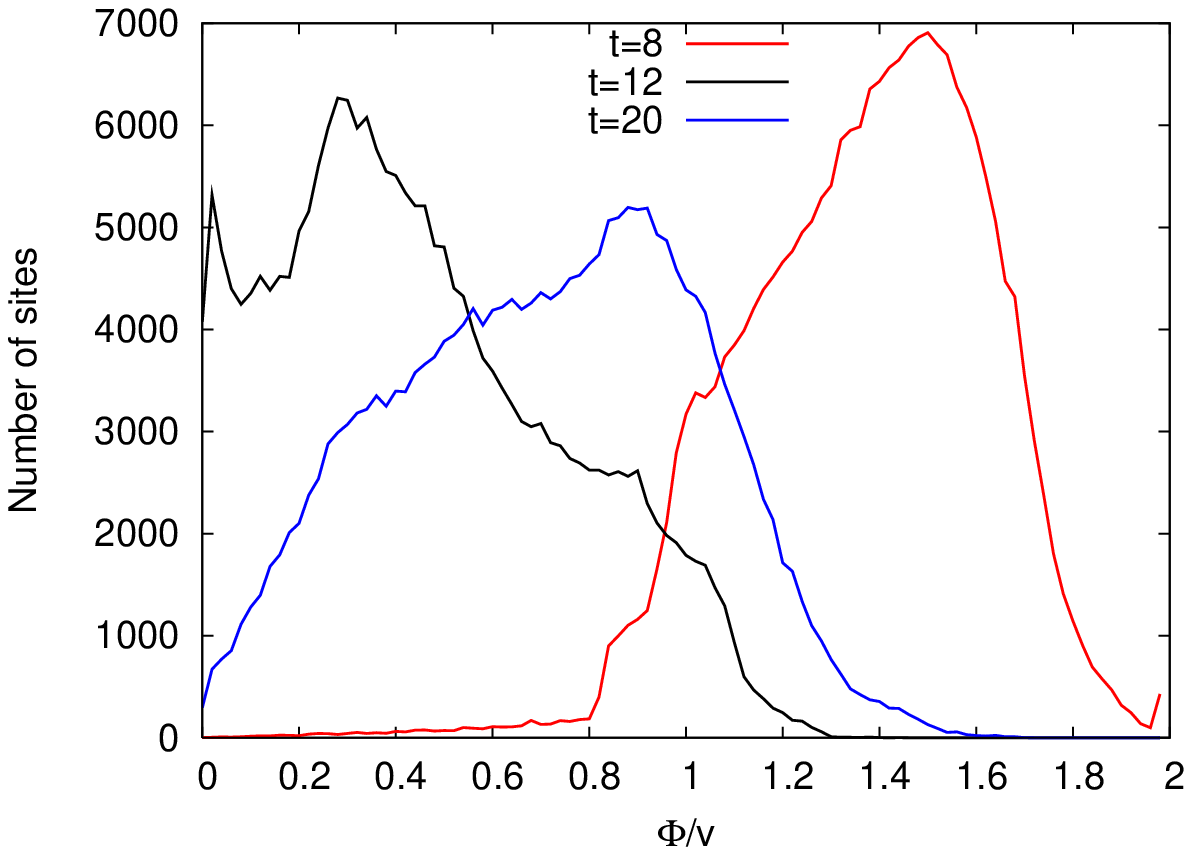}
\caption{Histograms that display the distribution of the Higgs length
  on the lattice at times $m_{\rm H} t = 8$, 12 and 20.}
\label{fig:hist}
}

\clearpage

\FIGURE{
\includegraphics[width=.49\textwidth,clip]{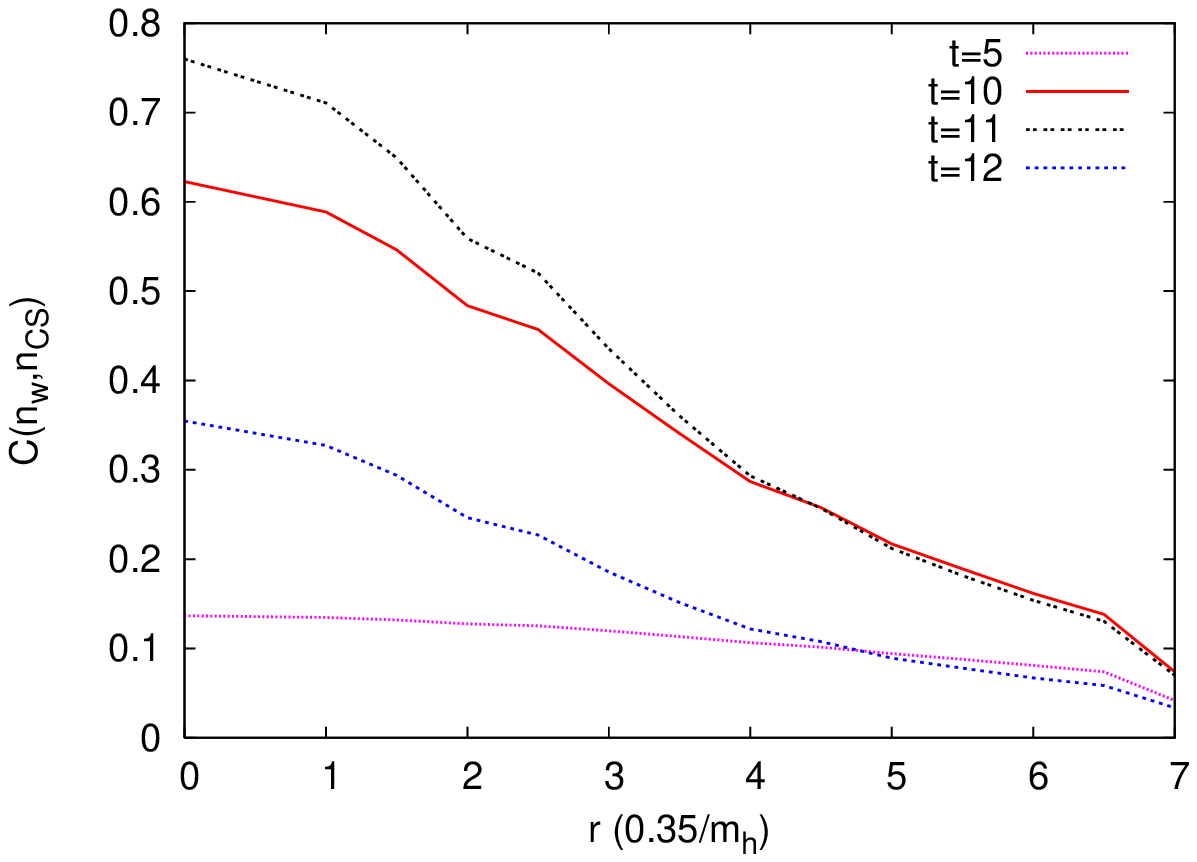}
\includegraphics[width=.49\textwidth,clip]{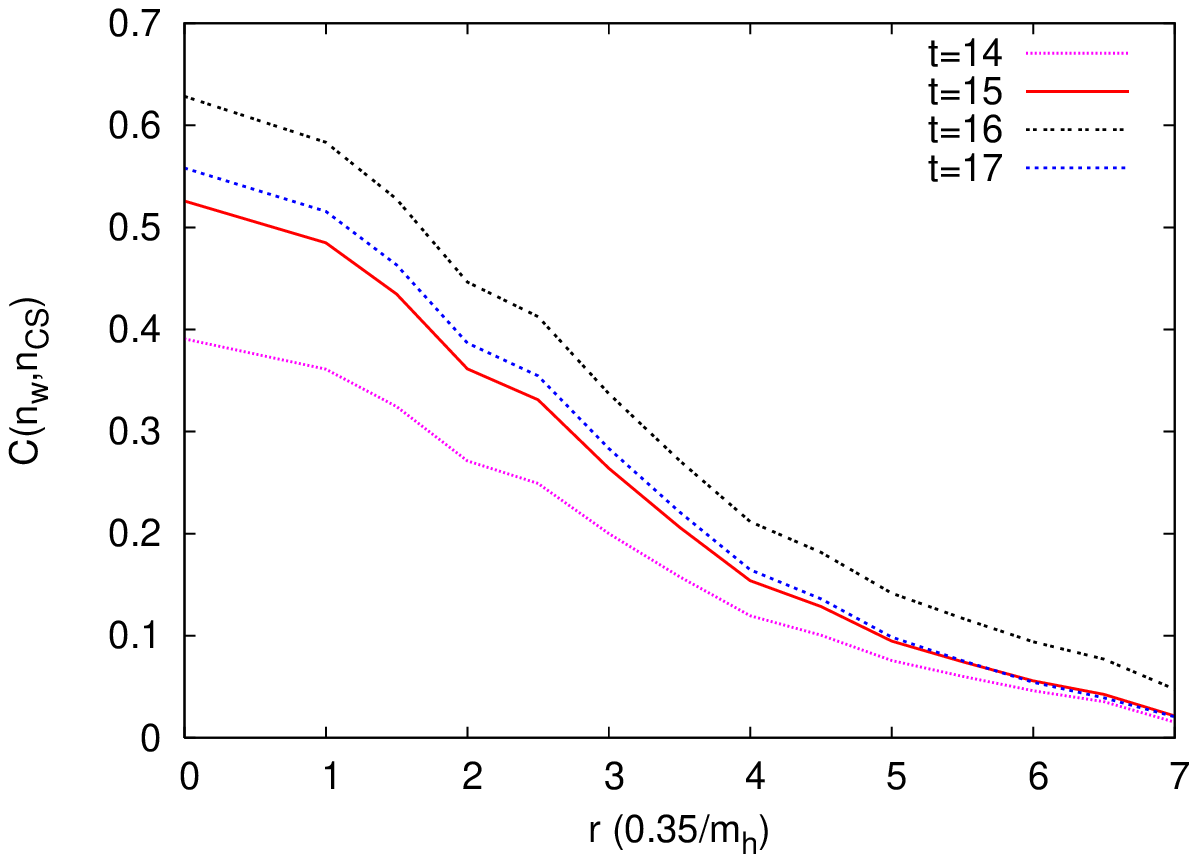}
\caption{The correlation $C(\vecr,t)$ between $\nncs$ and $\nnw$,
  defined in \eqref{wcscorrvsr},
  versus $r=|\vecr|$ at various times.}
\label{fig:wcscorrvsr}
}

\FIGURE{
\includegraphics[width=0.74\textwidth]{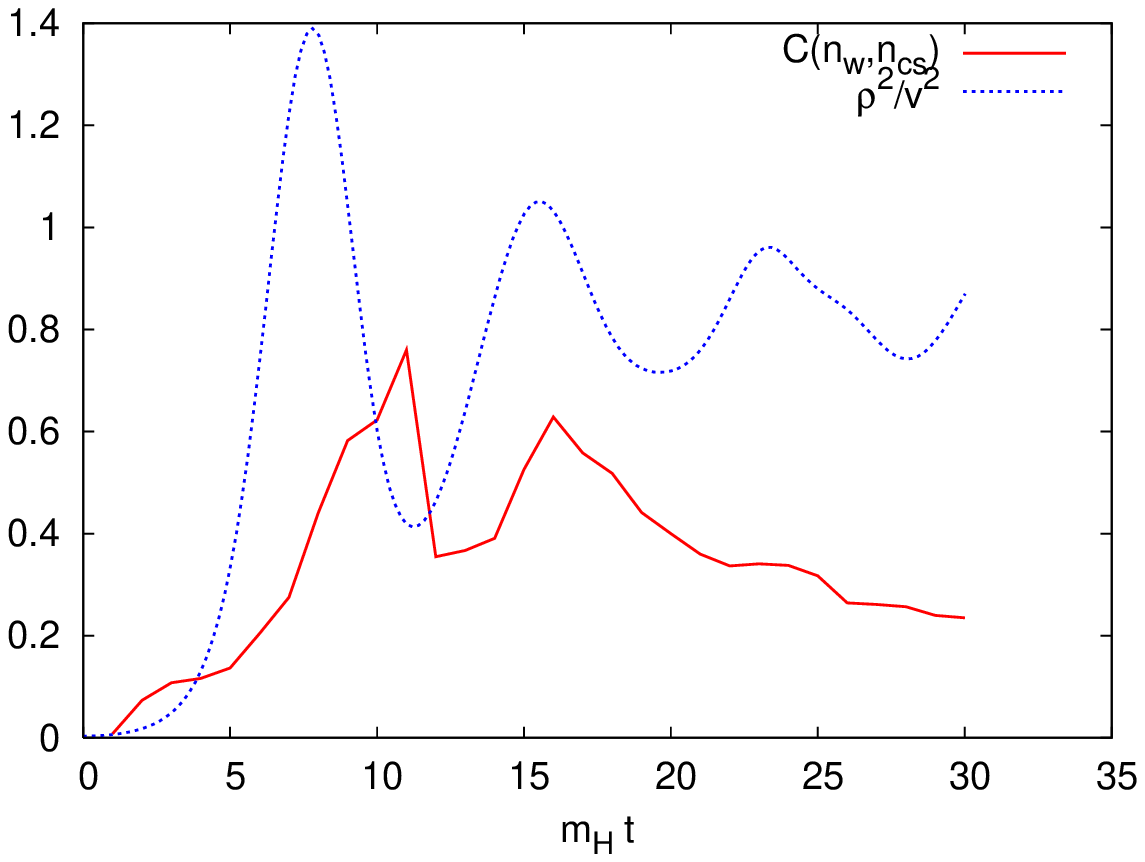}
\caption{The correlation
$C(0,t)$  versus time.
}
\label{fig:wcscorr}
}

\FIGURE{
\includegraphics[width=0.74\textwidth]{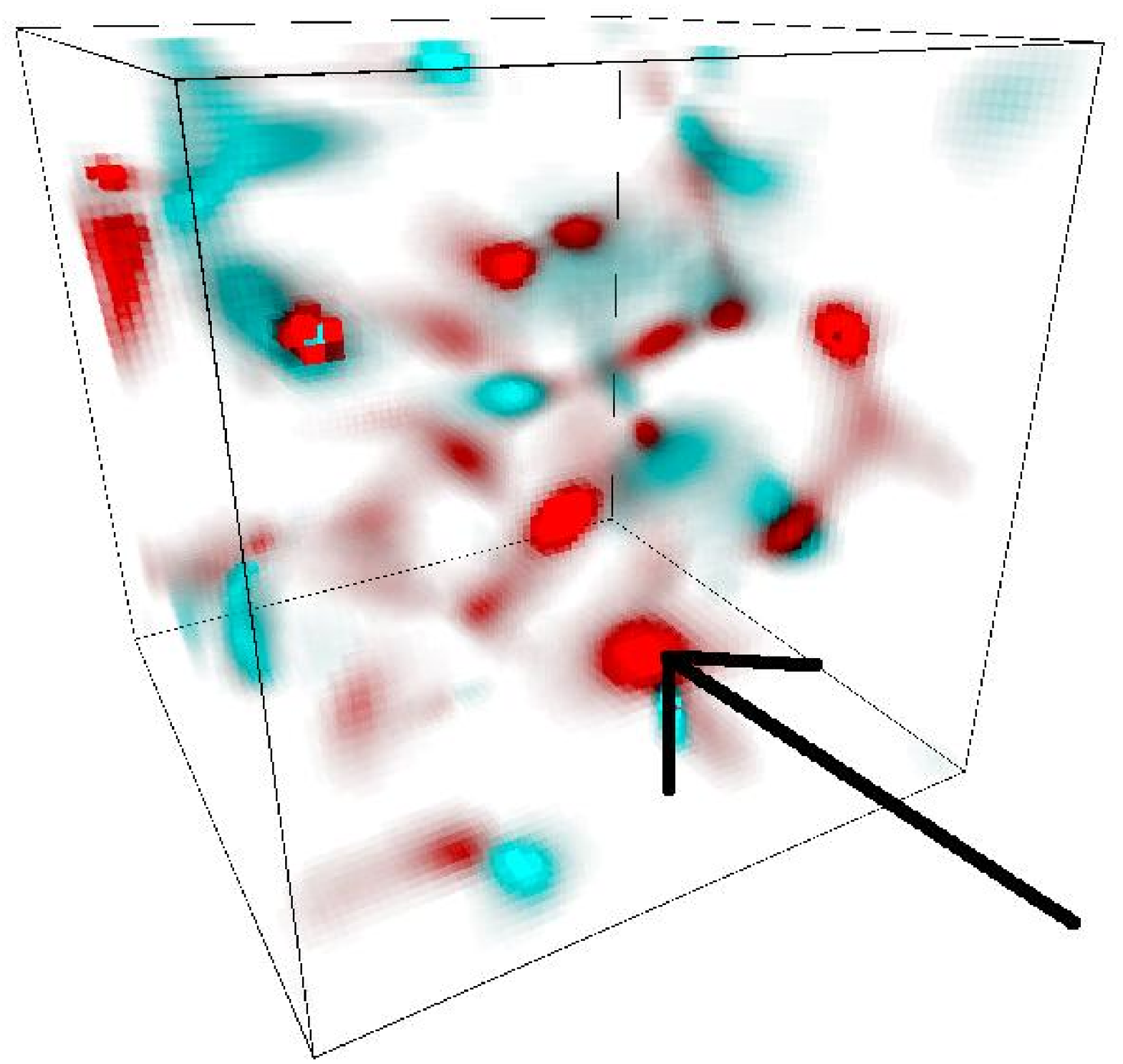}
\caption{
The winding number density at time $m_{\rm H} t = 1$ of the same
run as used before. The blob that we consider in this section is
indicated by the arrow.
  }
\label{fig:blob2}
}

\FIGURE{
\includegraphics[width=.49\textwidth]{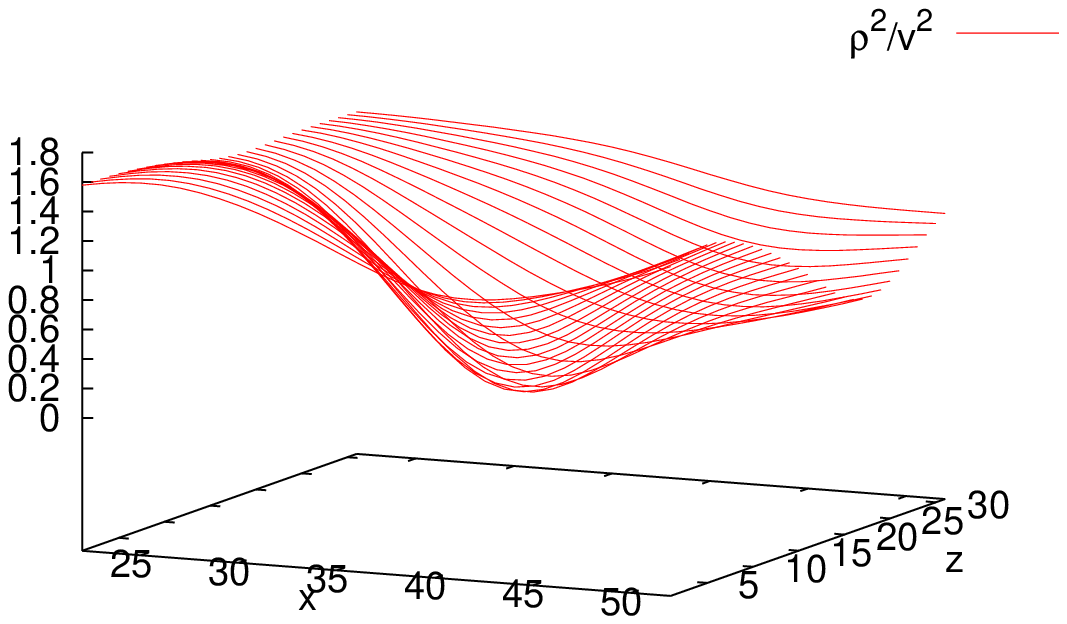}
\includegraphics[width=.49\textwidth]{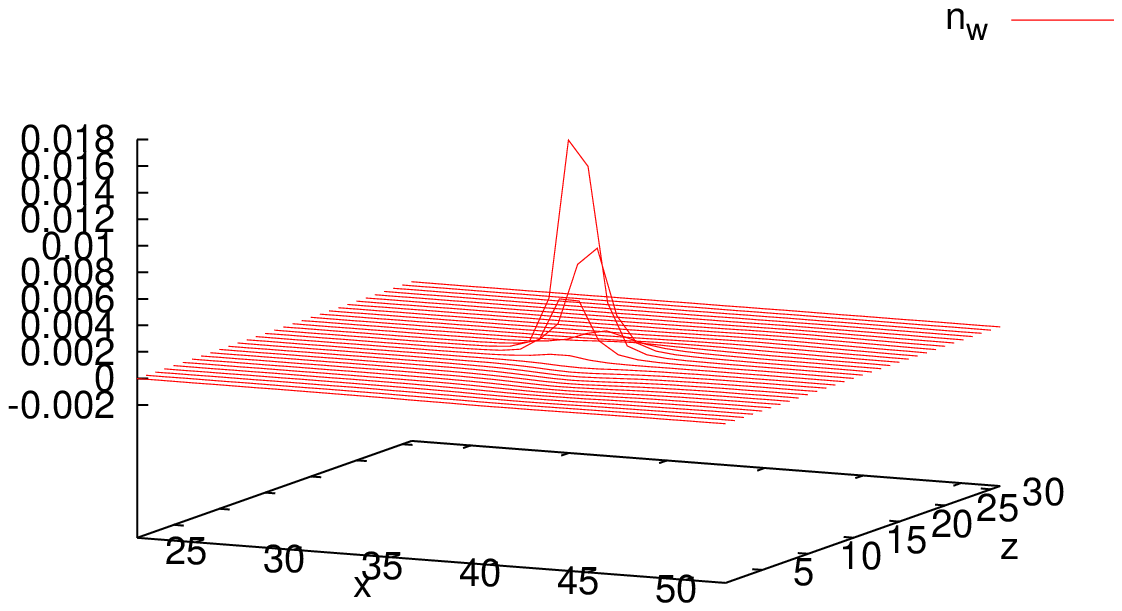}
\caption{Left the Higgs length (vertical) at time $m_{\rm H} t =2$ is
  plotted at the position of the blob, as function of the $x$ and $z$
  coordinates (a vertical slice). Right the winding number density at
  time $m_{\rm H} t = 2$ is plotted for the same slice through the blob.}
\label{fig:slice2}
}

\FIGURE{
\includegraphics[width=.49\textwidth,clip]{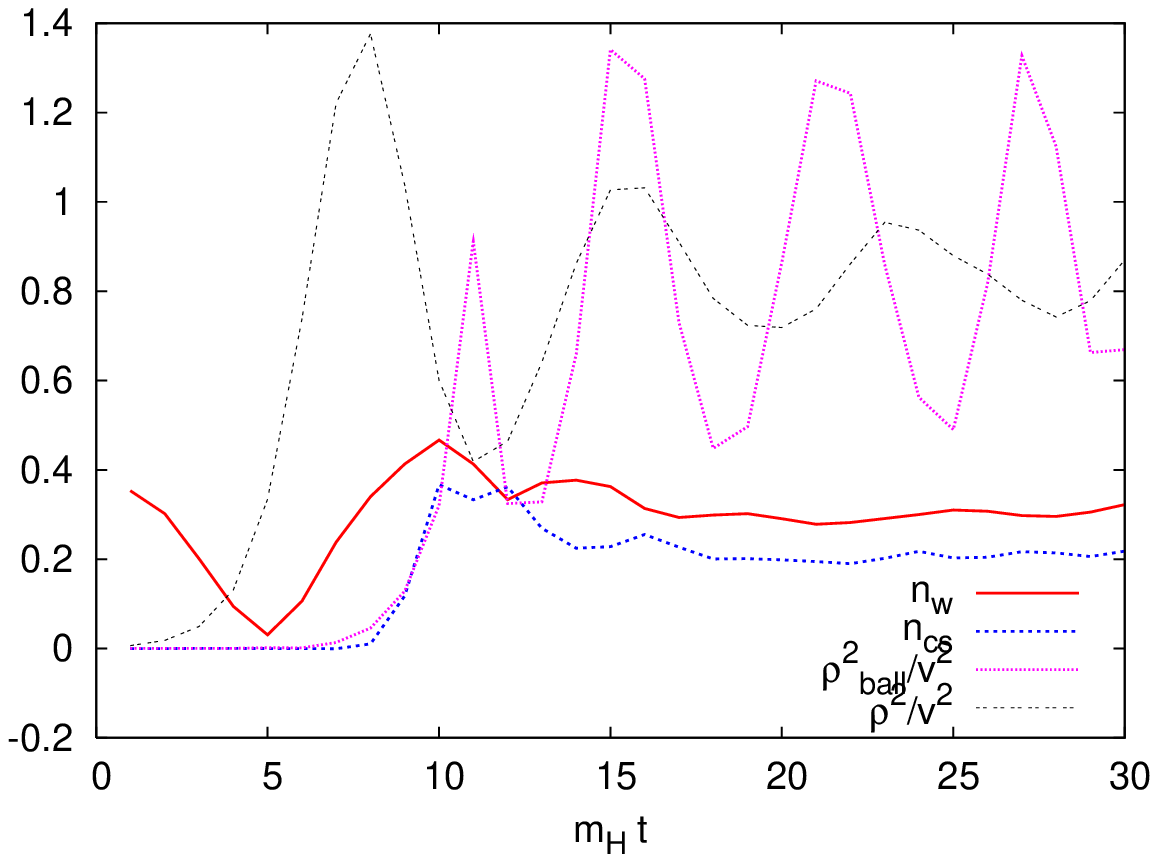}
\includegraphics[width=.49\textwidth,clip]{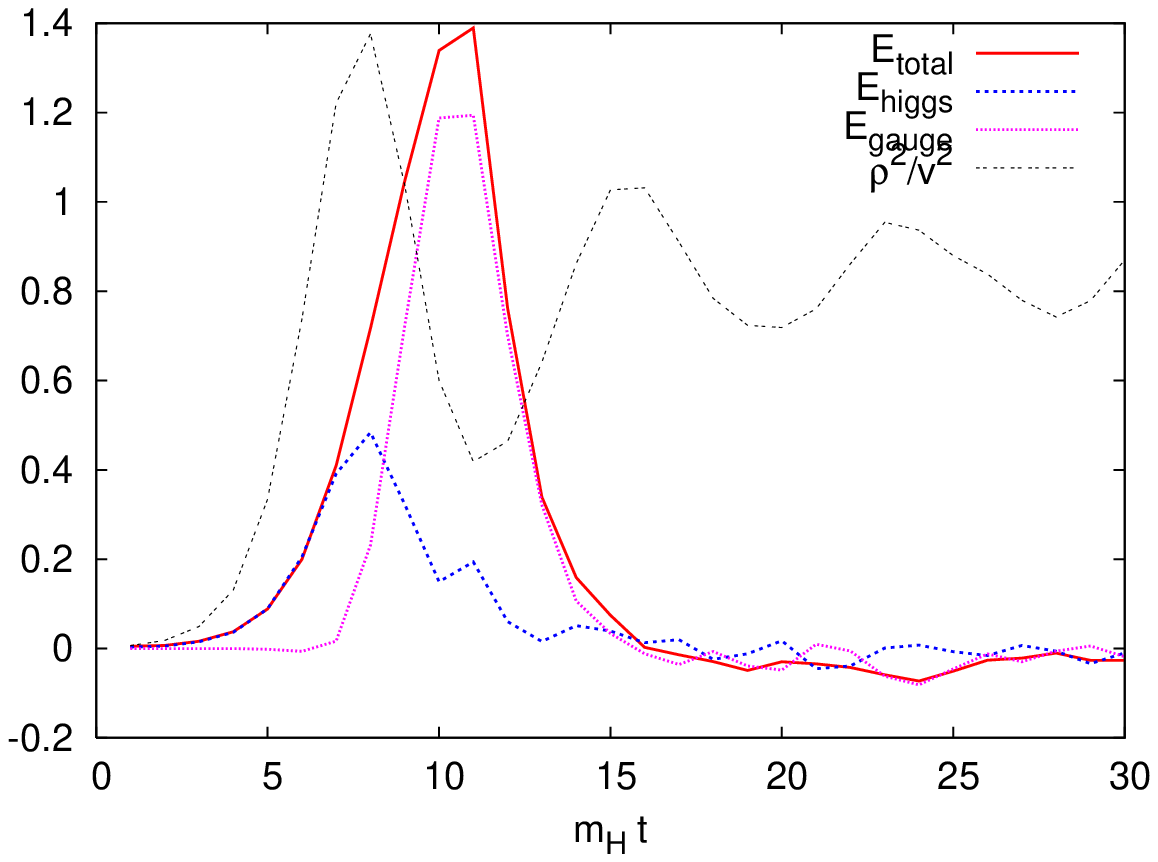}
\caption{
Left: $\nw^{\rm ball}$, $\NCS^{\rm ball}$ and
$\overline{\rh^2}^{\rm ball}$, for a ball with a radius of 6
lattice units ($2.1\, m_{\rm H}^{-1}$) around the center of the
blob. Right: excess energy,
$[\int_{\rm ball}d^3 x\,(\ep-\overline{\ep})]/E_{\rm sph}$,
in the same ball and its contributions from the Higgs field and
the gauge fields. }
\label{fig:ball}
}

\FIGURE{
\includegraphics[width=.49\textwidth]{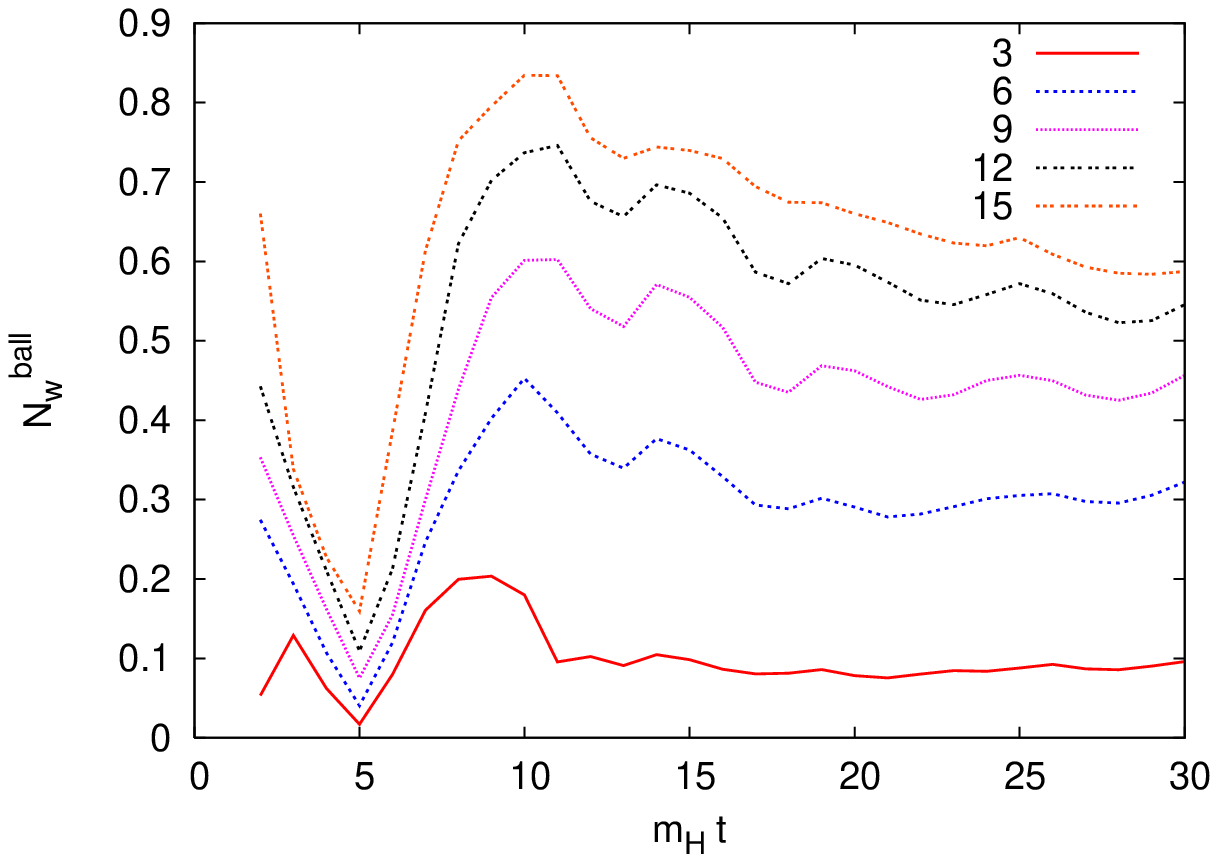}
\includegraphics[width=.49\textwidth]{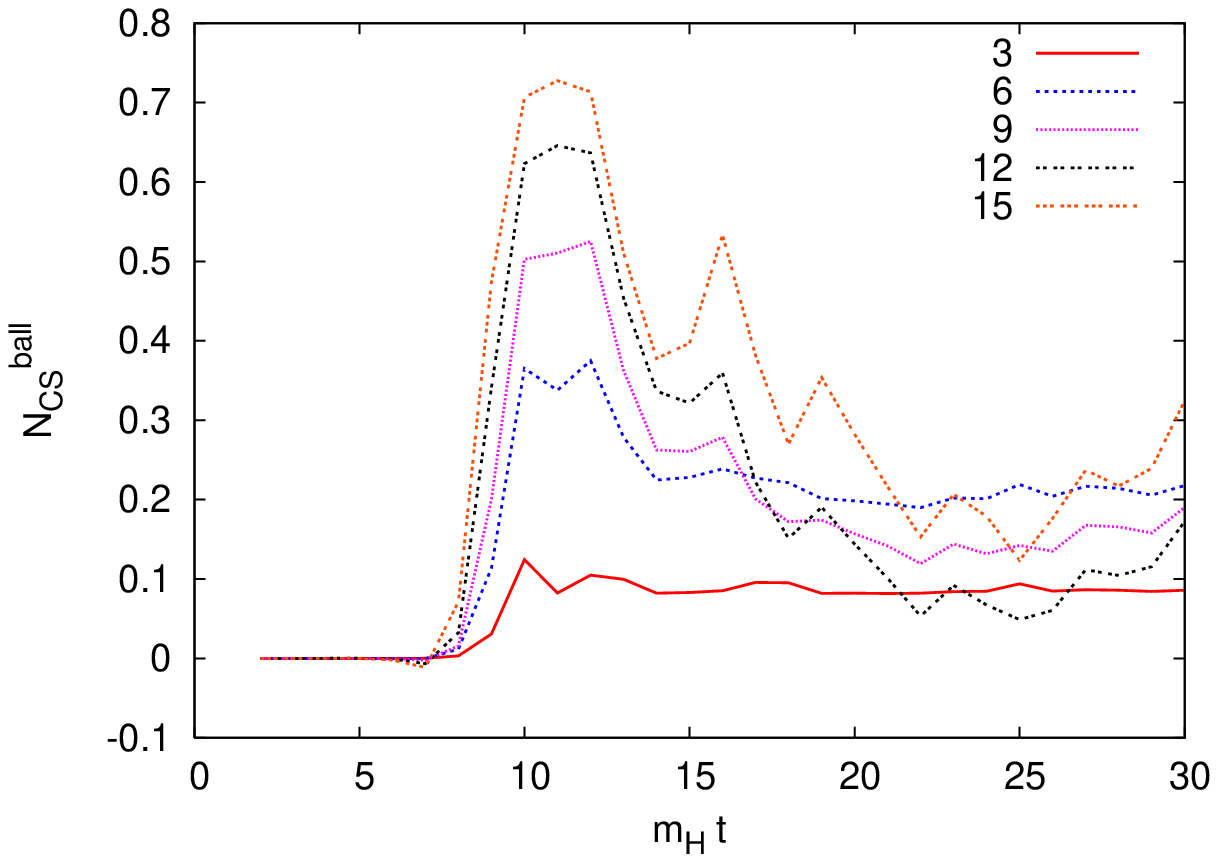}
\caption{
$\nw^{\rm ball}$ (left) and $\NCS^{\rm ball}$ (right)
for balls with varying radii, increasing
  from 3 lattice distances up to 15 lattice distances.}
\label{fig:varrad}
}

\FIGURE{
\includegraphics[width=0.74\textwidth]{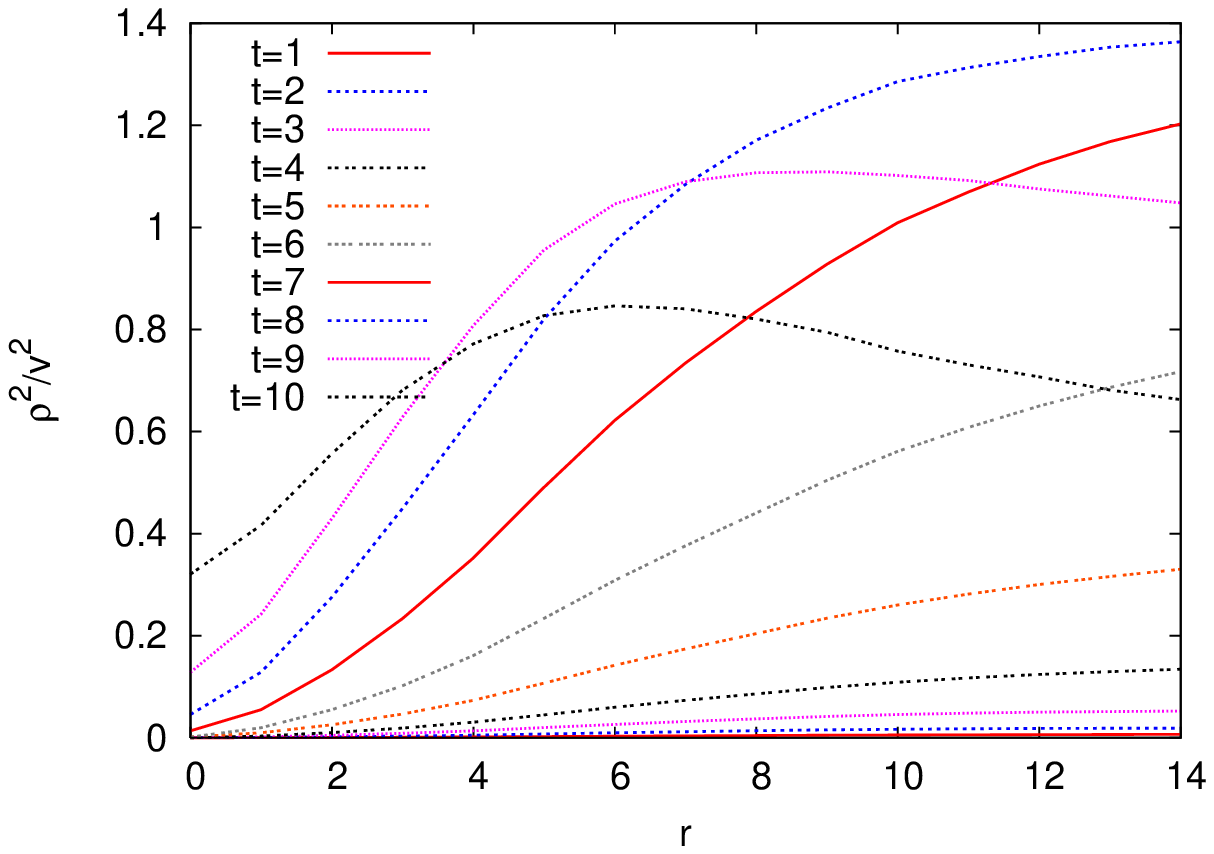}
\caption{Profiles of the normalized Higgs length
$\rh^2(r)/v^2$ for times
$m_{\rm H} t =1$ to $m_{\rm H} t=10$. On the horizontal axis
is the distance from
  the center, $r$, in lattice units (0.35 $m_{\rm H}^{-1}$).}
\label{fig:1hp}
}

\FIGURE{
\includegraphics[width=.49\textwidth]{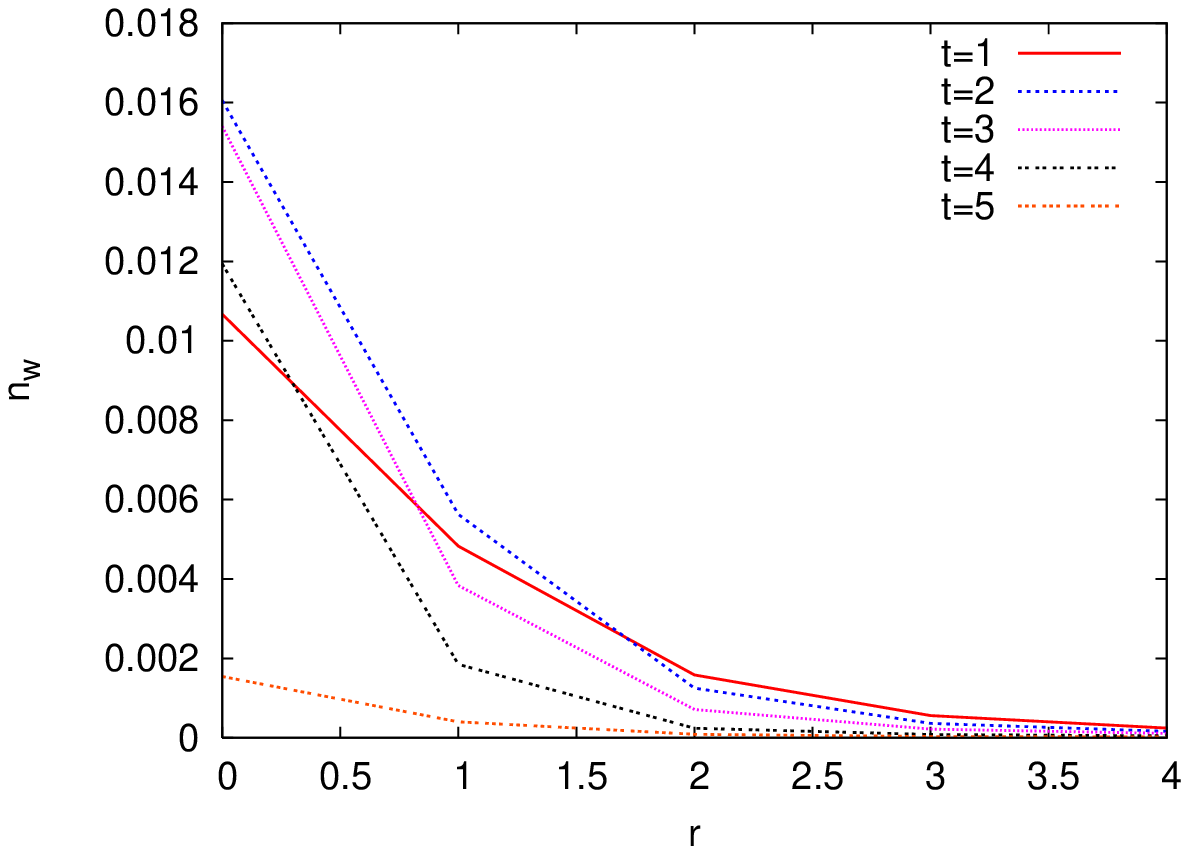}
\includegraphics[width=.49\textwidth]{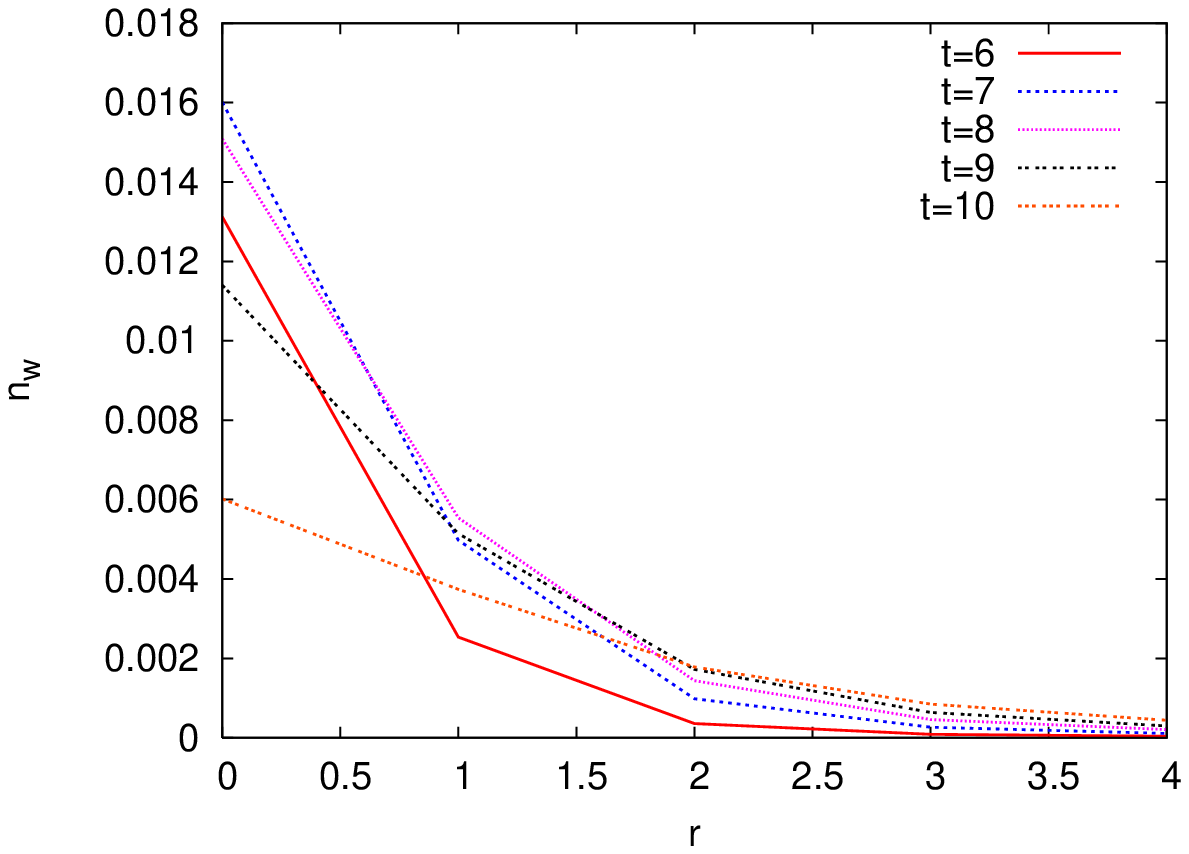}
\caption{
As in figure \ref{fig:1hp} for
$\nnw(r)$
for times $m_{\rm H} t =1$ to $m_{\rm H} t=5$ (left) and until
$m_{\rm H} t =10$ (right).
}
\label{fig:wp}
}

\FIGURE{
\includegraphics[width=0.74\textwidth]{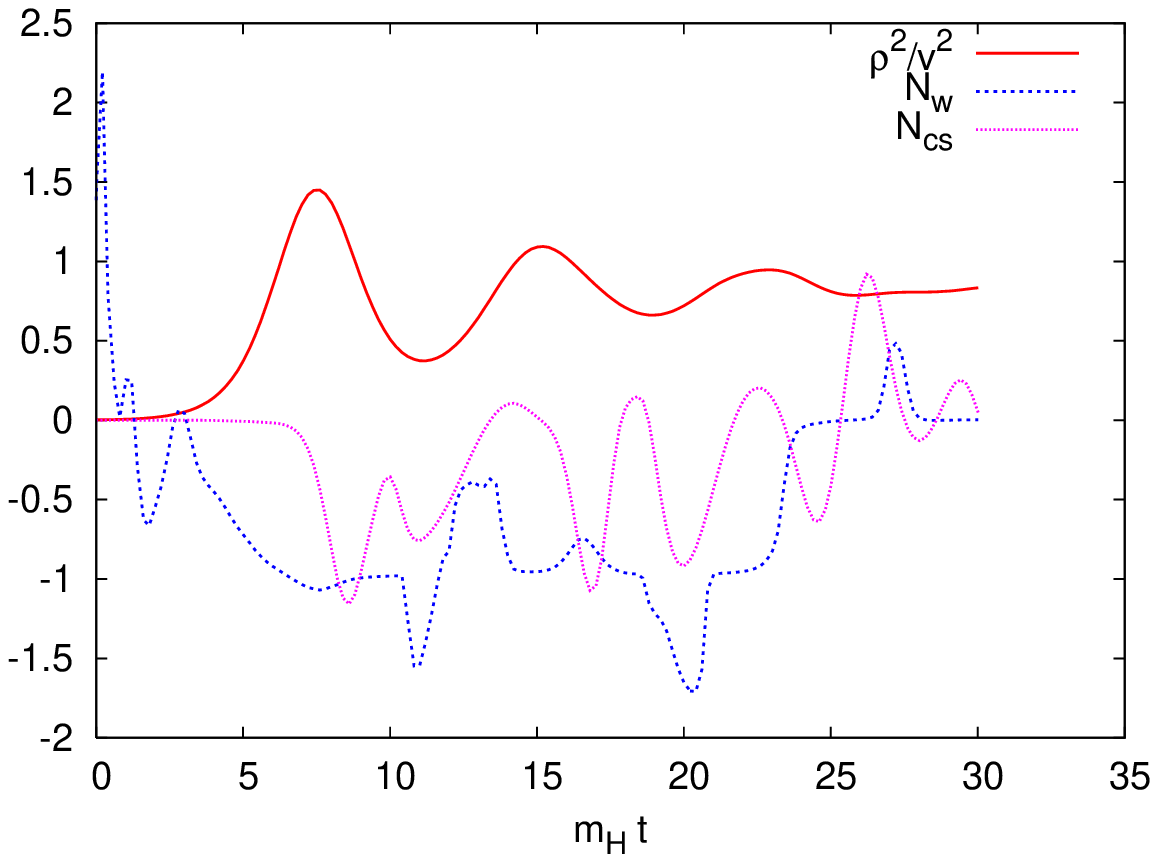}
\caption{
The analog of figure \ref{fig:gen30} for the trajectory with a
late transition,
$\overline{\rh^2}/v^2$,
total winding number $N_{\rm w}$ and total Chern-Simons number
$N_{\rm CS}$.}
\label{fig:gen31}
}

\FIGURE{
\includegraphics[width=0.74\textwidth]{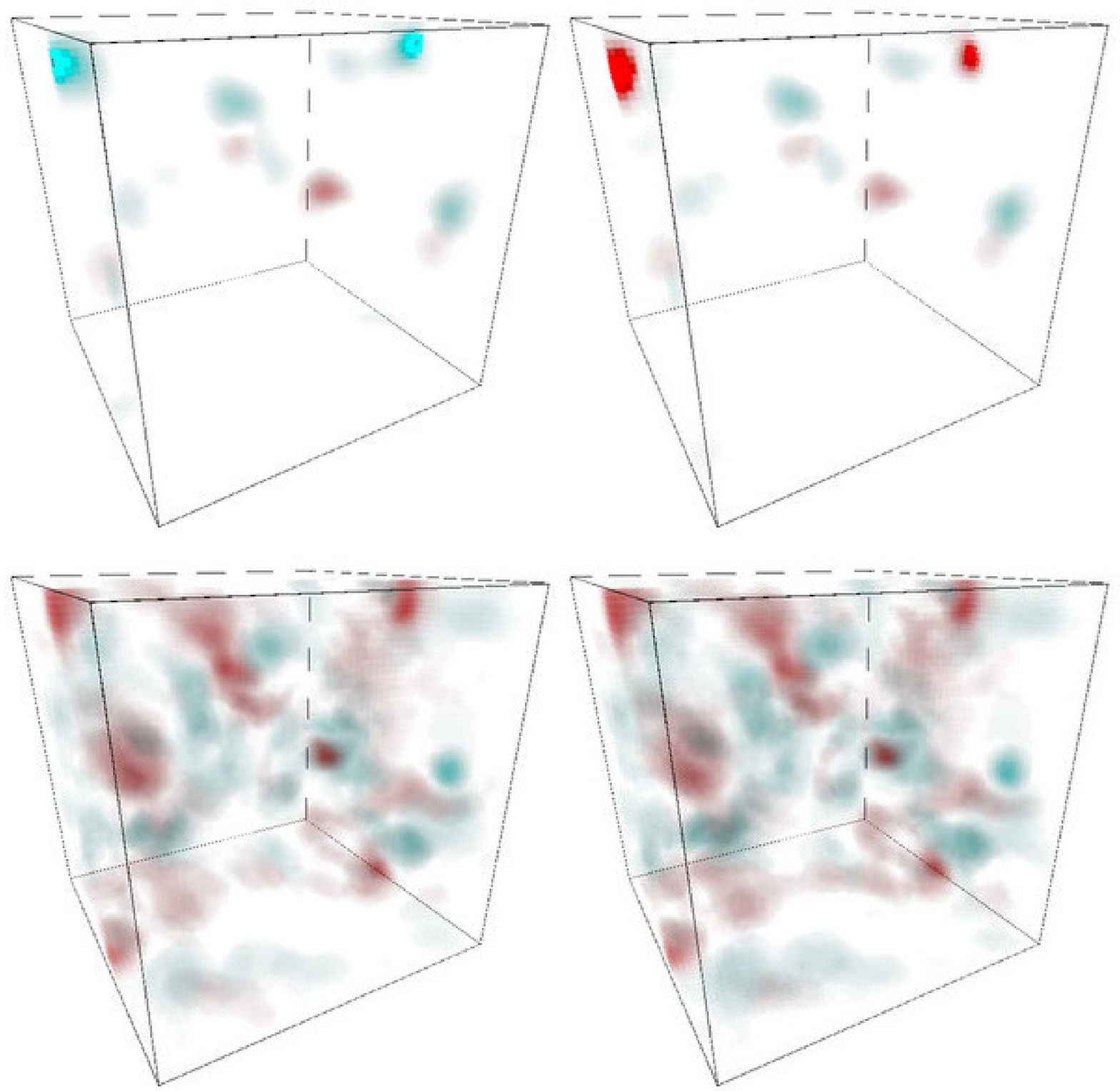}
\caption{
3D plot for the winding number-density (upper two figures) and
  Chern-Simons number-density (lower two figures)
  at times $m_{\rm H} t=23$ (left) and $m_{\rm H} t =24$
  (right).}
\label{fig:3drun31}
}

\FIGURE{
\includegraphics[width=\textwidth]{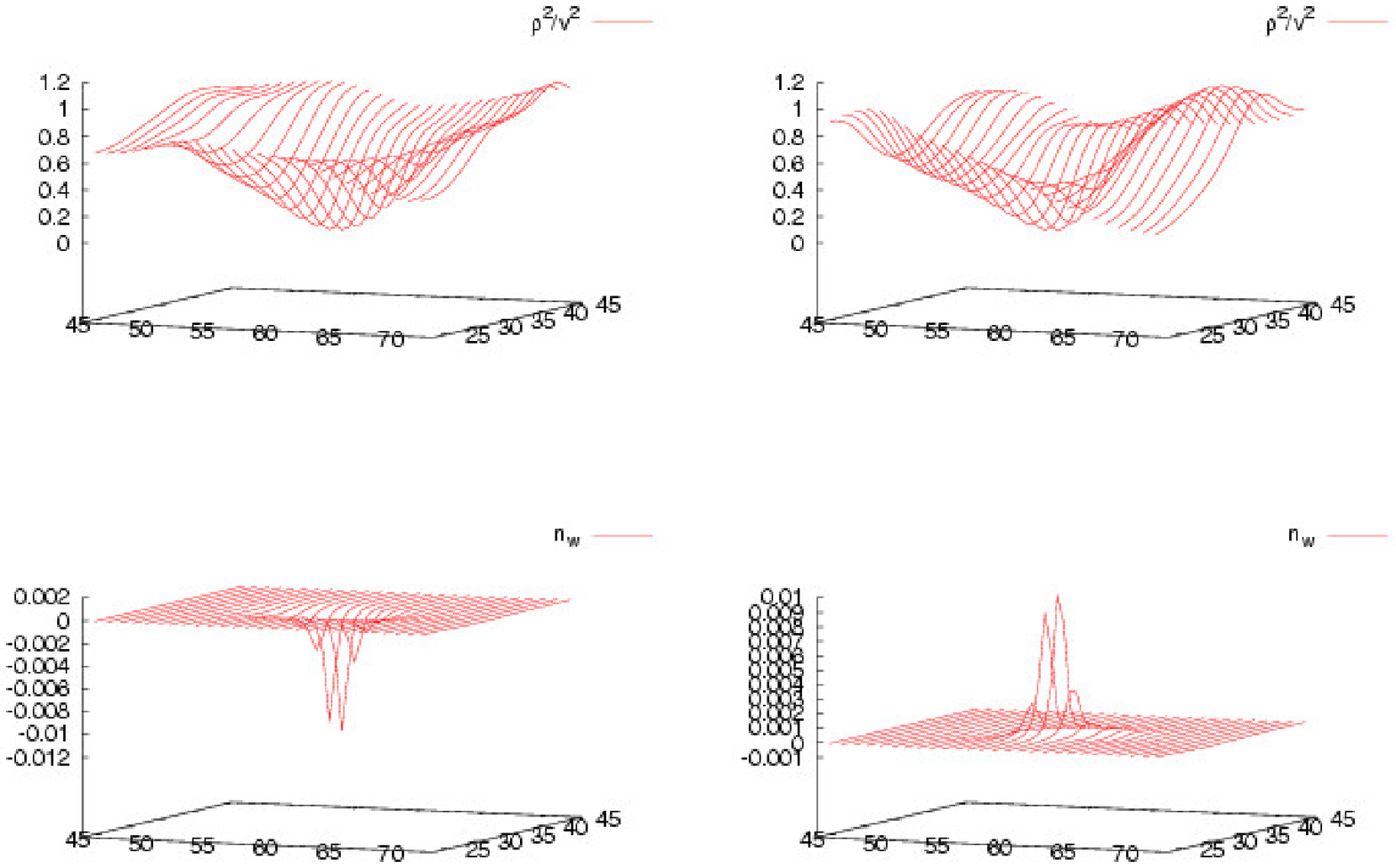}
\caption{
Upper plots: Higgs length
as function of the $x$ and $z$ coordinates through the blob. Left
is at time $m_{\rm H} t = 23$, right at time $m_{\rm H} t = 24$.
Lower plots: the corresponding winding number density.
}
\label{fig:nwdenslb}
}

\FIGURE{
\includegraphics[width=.49\textwidth]{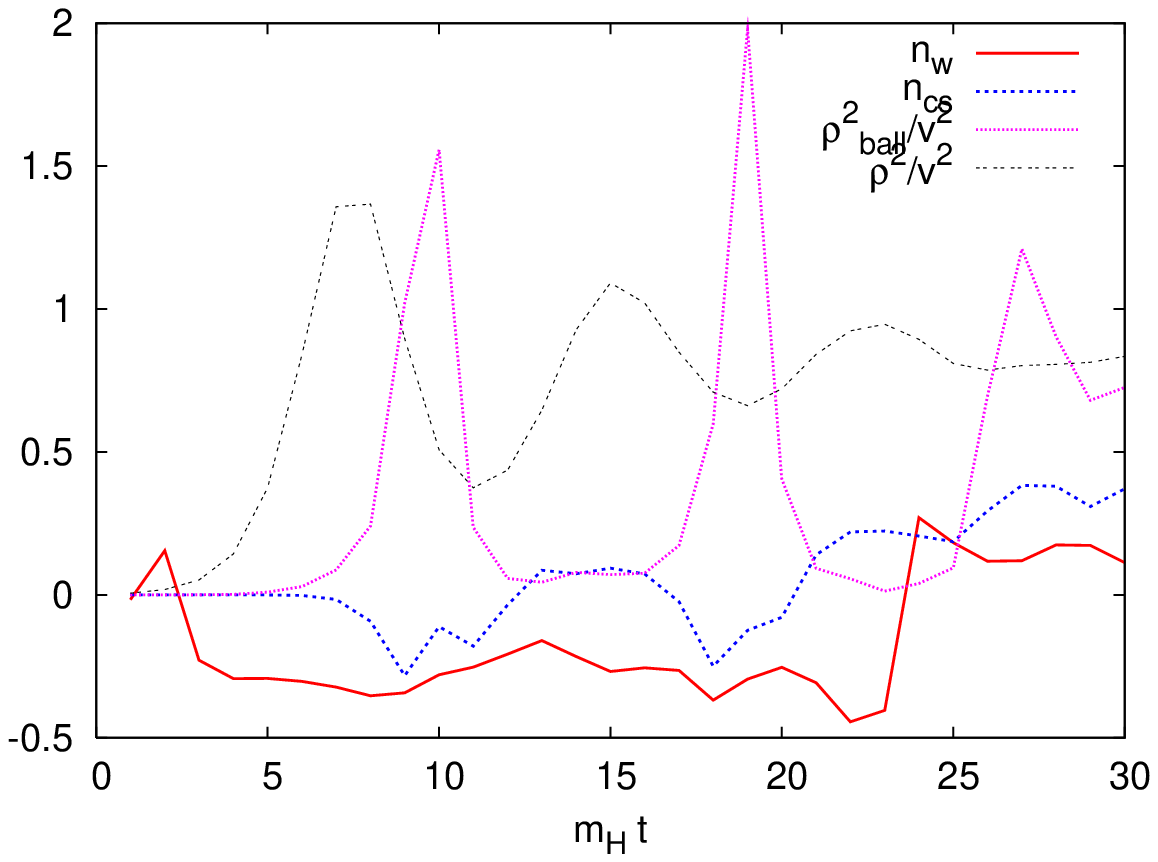}
\includegraphics[width=.49\textwidth]{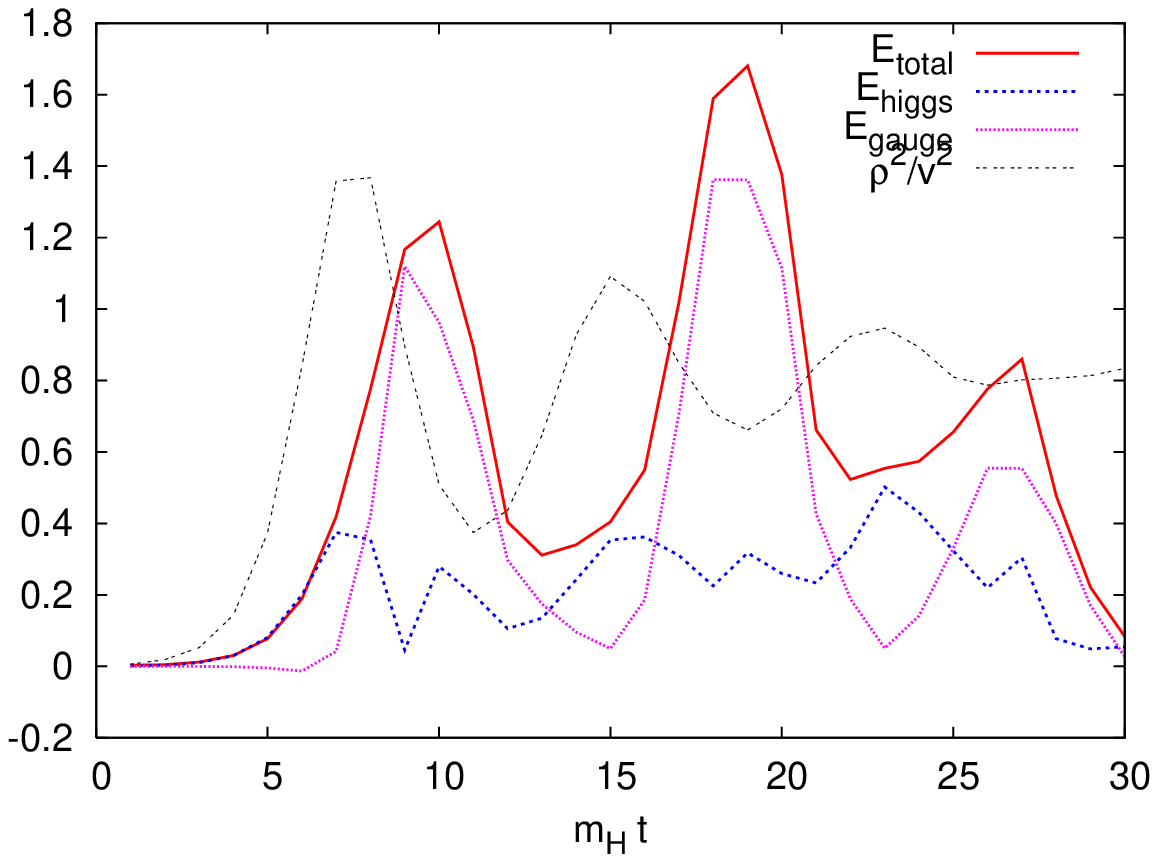}
\caption{
The analogue of figure \ref{fig:ball}: Left:
$\nw^{\rm ball}$, $\NCS^{\rm ball}$ and
$\overline{\rh^2}^{\rm ball}$, for a ball with a radius of 6
lattice units ($2.1\, m_{\rm H}^{-1}$) around the center of the
blob. Right: excess energy,
$[\int_{\rm ball}d^3 x\,(\ep-\overline{\ep})]/E_{\rm sph}$,
in the same ball and its contributions from the Higgs field and
the gauge fields. }
\label{fig:balll}
}

\FIGURE{
\includegraphics[width=.49\textwidth]{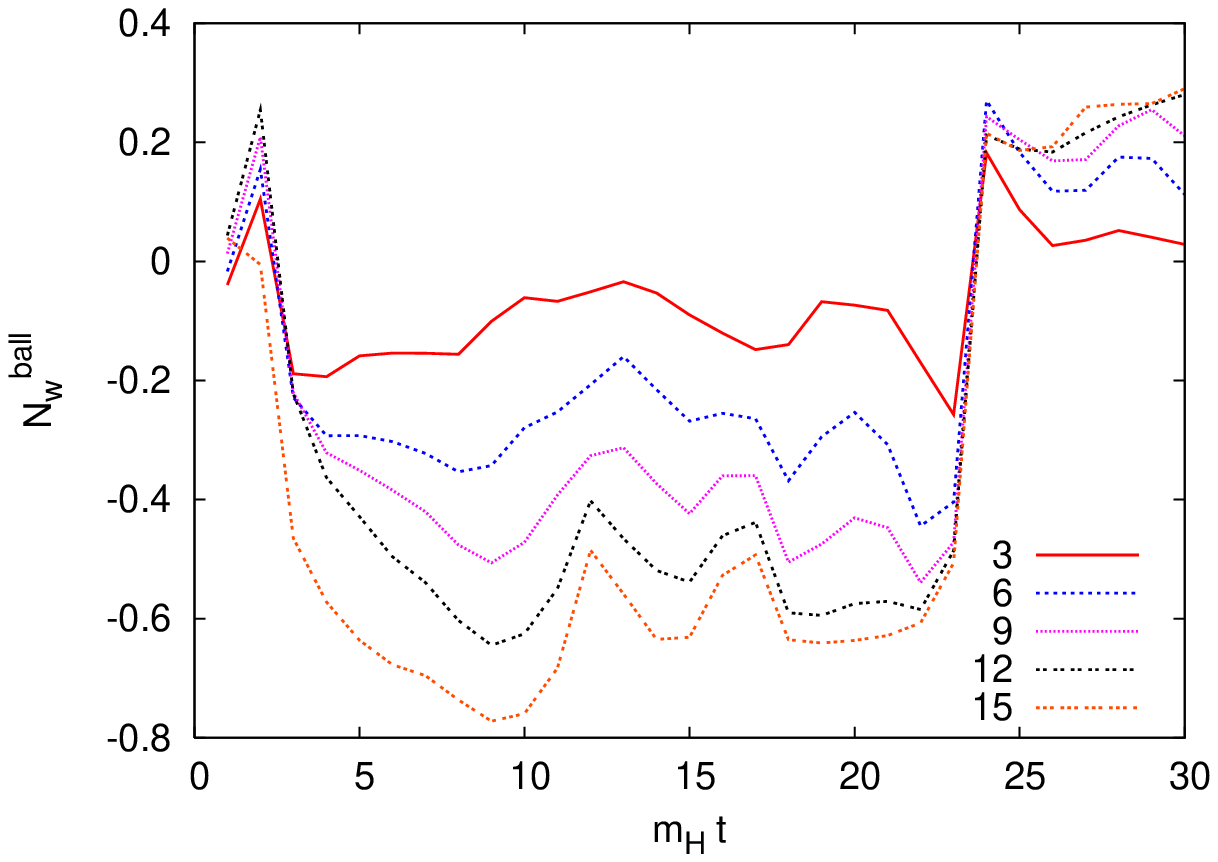}
\includegraphics[width=.49\textwidth]{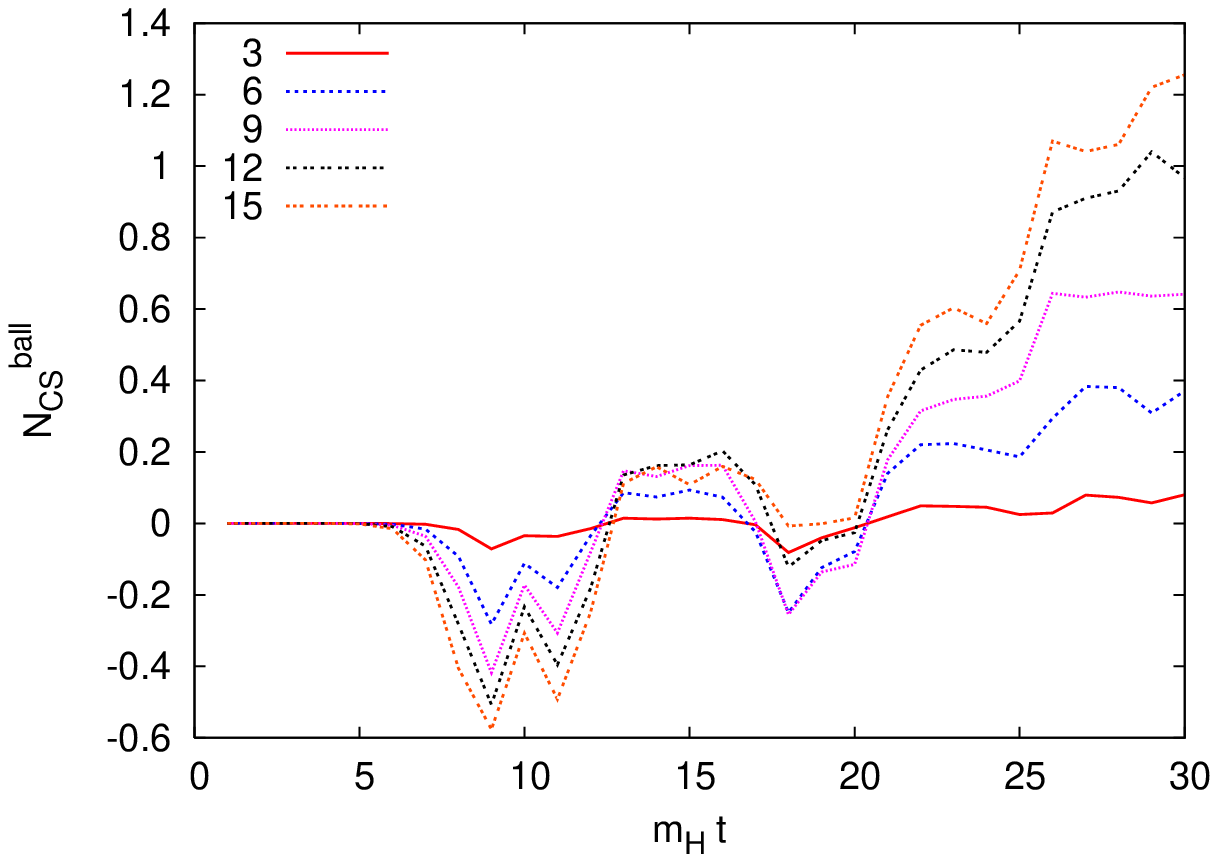}
\caption{
The analog of figure \ref{fig:varrad}: $\nw^{\rm ball}$ (left) and
$\NCS^{\rm ball}$ (right) for balls with varying radii, increasing
  from 3 to 15 lattice units.}
\label{fig:varradl}
}

\FIGURE{
\includegraphics[width=.49\textwidth]{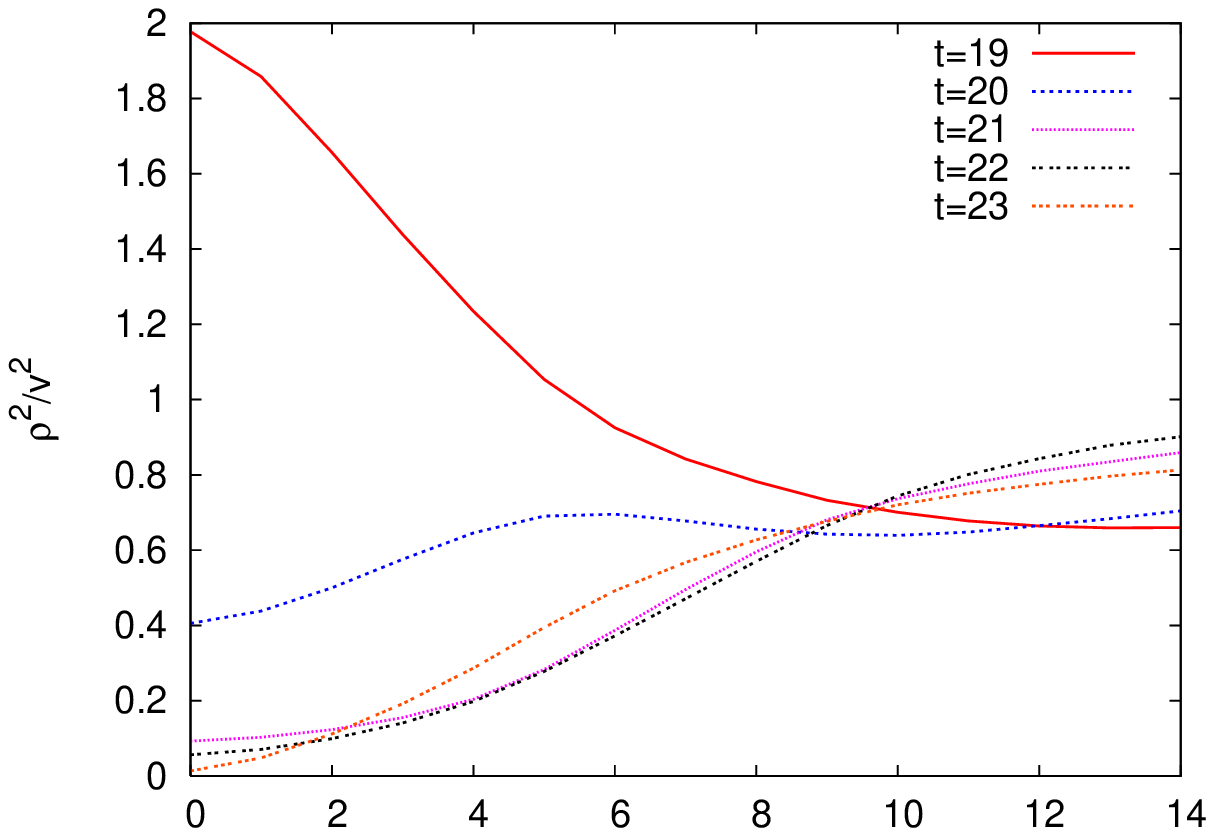}
\includegraphics[width=.49\textwidth]{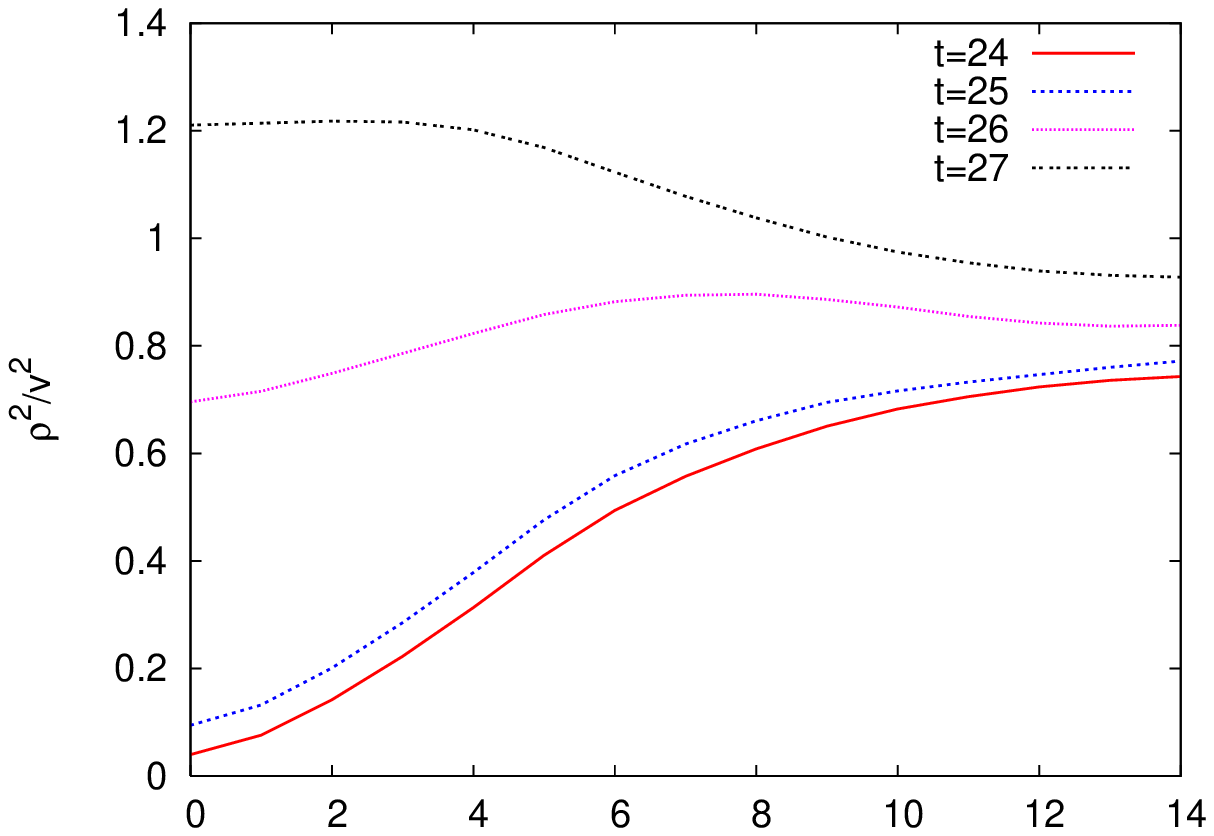}
\caption{
Profiles of the normalized Higgs length
$\rh^2(r)/v^2$
for times
$m_{\rm H} t =19$ to $m_{\rm H} t=23$ (left) and to $m_{\rm H}
t=27$ (right). On the horizontal axis is the distance from the
center $r$ in lattice units (0.35 $m_{\rm H}^{-1}$).}
\label{fig:latehp}
}

\FIGURE{
\includegraphics[width=.49\textwidth]{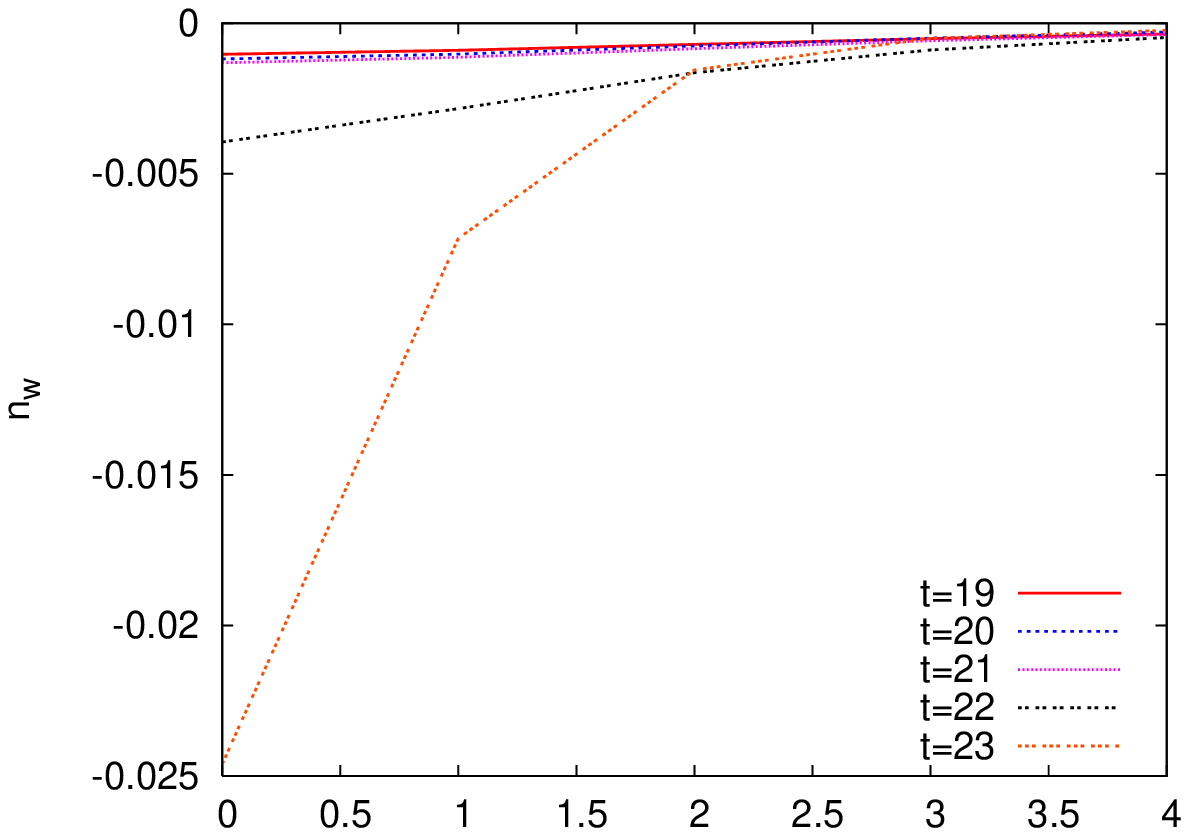}
\includegraphics[width=.49\textwidth]{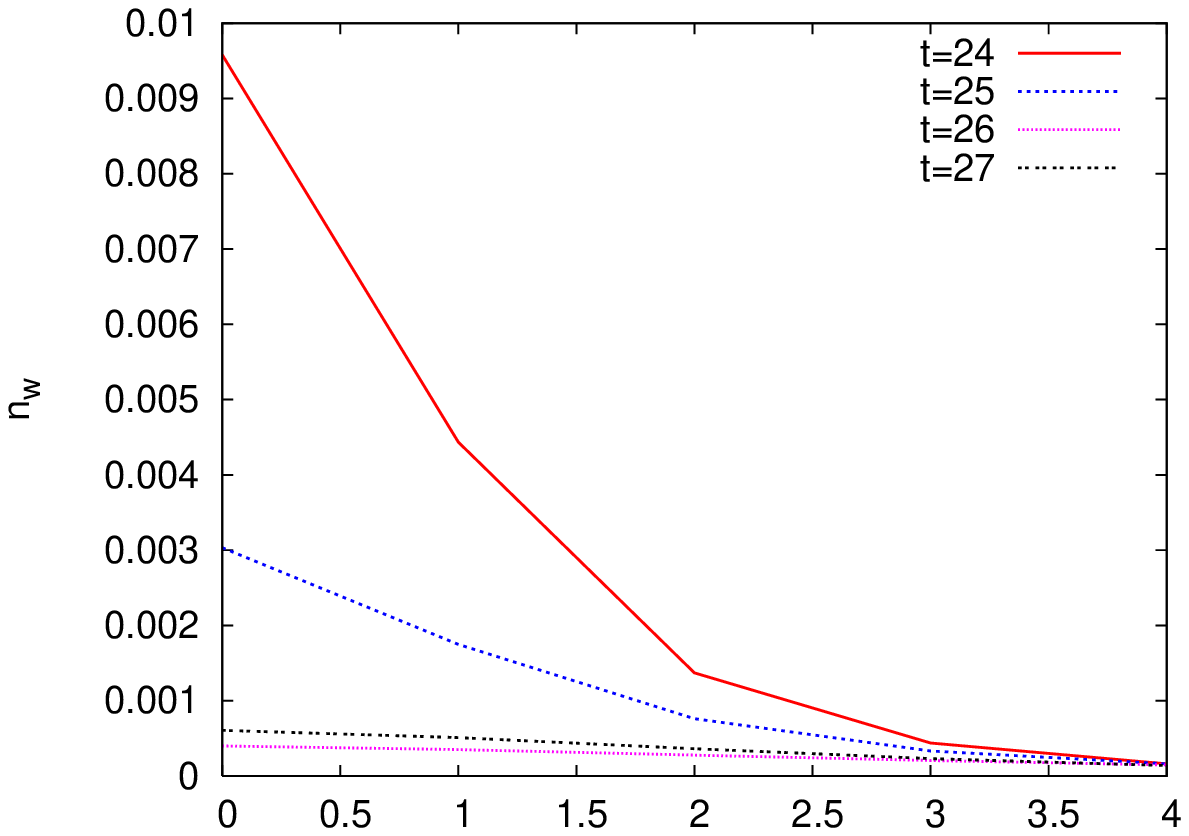}
\caption{
As in figure \ref{fig:latehp} for $\nnw(r)$.}
\label{fig:latewp}
}
\FIGURE{
\includegraphics[width=0.74\textwidth,clip]{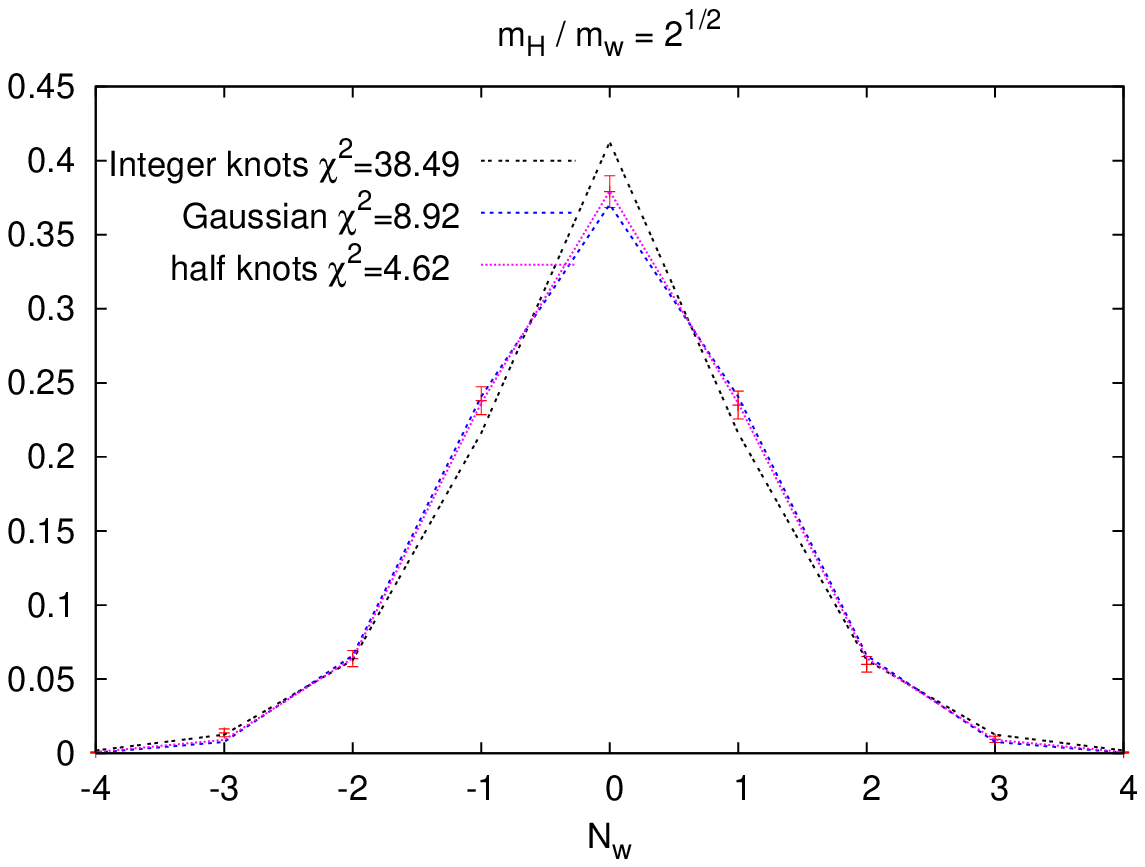}
\includegraphics[width=0.74\textwidth,clip]{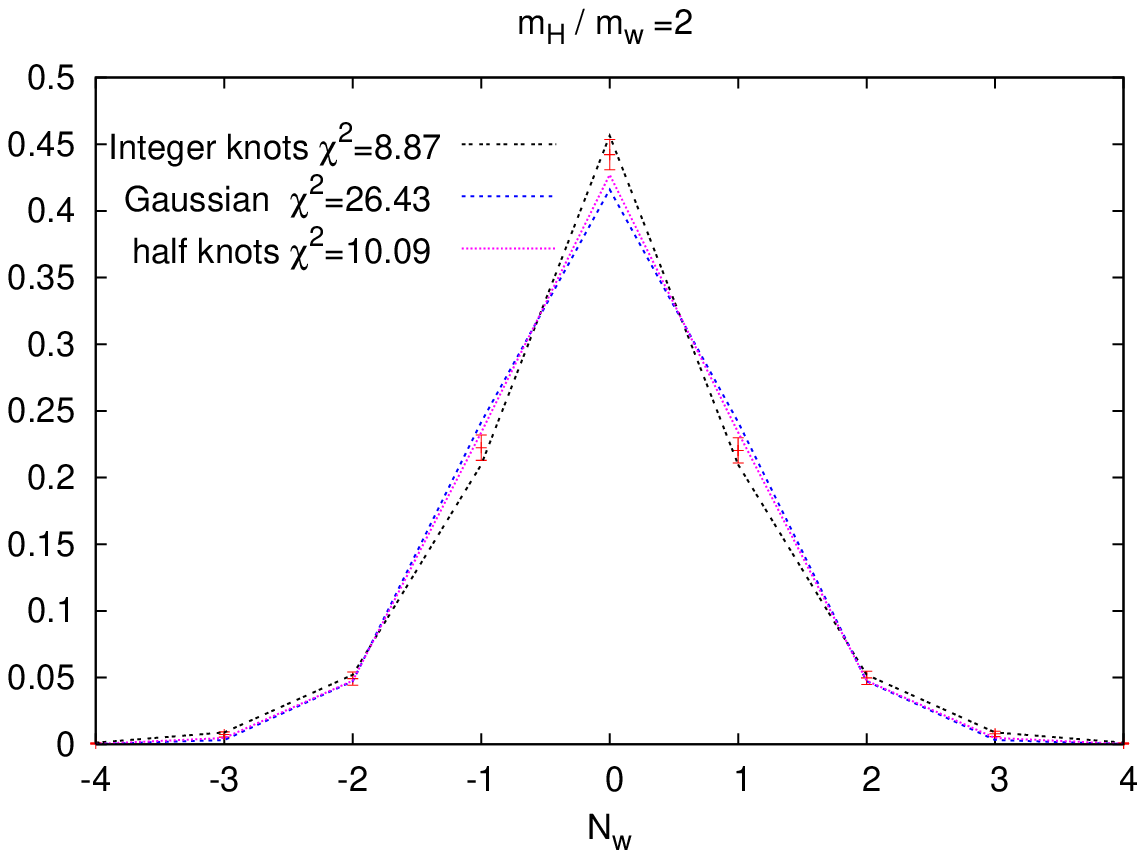}
\caption{Normalized distribution of winding numbers at $\mh t = 30$ for
$\mh = \sqrt{2}\, m_W$ and $\mh = 2\, m_W$.}
\label{fig:distnw}
}

\FIGURE{
\includegraphics[width=0.74\textwidth,clip]{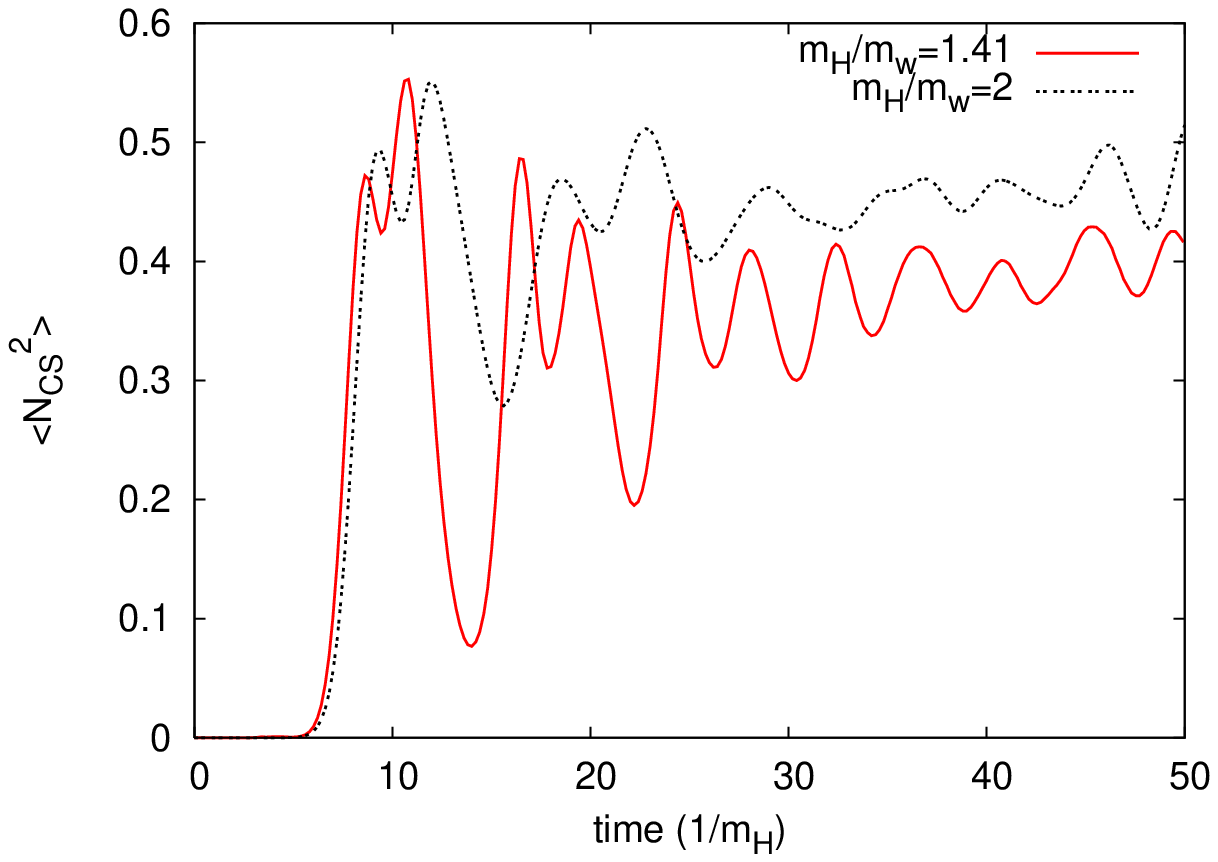}

\caption{Chern-Simons susceptibility $\langle\NCS^2(t)\rangle$
for $\mh = \sqrt{2}\, m_W$ and $\mh = 2\, m_W$.}
\label{fig:suscept}
}

\end{document}